\documentclass[usenatbib,onecolumn]{mn2e}
\usepackage{natbib}
\usepackage{epsfig}
\usepackage{lscape}
\usepackage{psfig}
\usepackage{amsmath,amssymb}
\usepackage{color}

\catcode`\@=11
\def\gta{\ifmmode{\,\mathrel{\mathpalette\@versim>\,}}
    \else{$\,\mathrel{\mathpalette\@versim>}\,$}\fi}
\def\lta{\ifmmode{\,\mathrel{\mathpalette\@versim<\,}}
    \else{$\,\mathrel{\mathpalette\@versim<}\,$}\fi}
\def\@versim#1#2{\lower 2.9truept \vbox{\baselineskip 0pt \lineskip
    0.5truept \ialign{$\m@th#1\hfil##\hfil$\crcr#2\crcr\sim\crcr}}}
\catcode`\@=12  
\def\ArhoR{A_{\rho{R}}}
\newcommand\Arhoz{A_{\rho{z}}}
\newcommand\ApR{A_{pR}}
\newcommand\Apz{A_{pz}}
\newcommand\cmminusone{\,{\rm cm}^{-1}}
\newcommand\czero{c_0}

\renewcommand\d{{\rm d}}
\newcommand\Dm{\mathcal{D}}
\newcommand\erg{\,{\rm erg}}

\newcommand\fzero{{f_0}}
\newcommand\grad{{\nabla}}

\renewcommand\i{{\rm i}}

\newcommand\kz{k_z}
\newcommand\kR{k_R}

\newcommand\kphi{k_{\phi}}
\newcommand\kv{{\bf k}}
\newcommand\kzksq{{\kz^2\over k^2}}

\newcommand\pd{\partial}
\newcommand\pzero{{p_0}}
\newcommand\sminusone{\,{\rm s}^{-1}}
\newcommand\szero{{s_0}}
\newcommand\Tzero{T_0}
\newcommand\vz{{v_z}}
\newcommand\vR{{v_R}}
\newcommand\vphi{v_{\phi}}

\newcommand\vvzero{\vv_0}
\newcommand\vzerophi{v_{0\phi}}
\newcommand\vzeroR{v_{0R}}
\newcommand\vzeroz{v_{0z}}

\newcommand\vv{{\bf v}}

\newcommand\Deltaone{\Delta_1}
\newcommand\Deltatwo{\Delta_2}
\newcommand\Deltathree{\Delta_3}
\newcommand\Deltafour{\Delta_4}
\newcommand\Deltafive{\Delta_5}
\newcommand\rhozero{\rho_0}

\newcommand\omegaA{\omega_{\rm A}}
\newcommand\omegaAsq{\omega^2_{\rm A}}

\newcommand\omegad{\omega_{\rm d}}

\newcommand\omegadsq{\omega_{\rm d}^2}
\newcommand\omegaBV{\omega_{\rm BV}}
\newcommand\omegaBVsq{\omega^2_{\rm BV}}

\newcommand\omegarotsq{\omega^2_{\rm rot}}
\newcommand\omegacmag{\omega_{\rm c,mag}}
\newcommand\omegacphisq{\omega^2_{\rm c,\phi}}
\newcommand\omegacphi{\omega_{\rm c,\phi}}

\newcommand\omegac{\omega_{\rm c}}

\newcommand\Omegaz{\Omega_{z}}
\newcommand\OmegaR{\Omega_{R}}
\newcommand\omegath{\omega_{\rm th}}

\newcommand\kelvin{\,{\rm K}} 

\newcommand\bv{{\bf b}}
\newcommand\de{\partial}

\newcommand\Bv{{\bf B}}
\newcommand\Bvzero{\Bv_0}
\newcommand\Bzero{B_0}
\newcommand\Bzerophi{B_{0\phi}}
\newcommand\BzeroR{B_{0R}}
\newcommand\Bzeroz{B_{0z}}
\newcommand\bzeroR{b_{0R}}
\newcommand\bzeroz{b_{0z}}
\newcommand\bzerophi{b_{0\phi}}
\newcommand\Bphi{B_{\phi}}
\newcommand\BR{B_{R}}
\newcommand\Bz{B_{z}}

\newcommand\Qv{{\bf Q}}
\newcommand\ds{\displaystyle}
\renewcommand\div{\nabla \cdot}
\newcommand\rot{\nabla \times}
\newcommand\e{{\rm e}}

\newcommand\bvzero{\bv_0}
\newcommand\vvA{\vv_{\rm A}}

\renewcommand\S{\mathcal{S}}
\renewcommand\L{\mathcal{L}}
\newcommand\LT{\mathcal{L}_T}
\newcommand\Lrho{\mathcal{L}_{\rho}}
\newcommand\omegaca{\omega_{\rm c,a}}
\newcommand\omegacasq{\omega^2_{\rm c,a}}

\defcitealias{Nip10}{N10}

\begin{document}

\date{Accepted 2012 September 24. Received 2012 September 24; in original form 2012 June 18}
\title[Stability of a rotating plasma]{Thermal stability of a weakly magnetized rotating plasma} 
\author[C. Nipoti and L. Posti]{Carlo Nipoti\thanks{E-mail: carlo.nipoti@unibo.it} and Lorenzo Posti\thanks{E-mail: lorenzo.posti@gmail.com}
\\
Dipartimento di Astronomia, Universit\`a di Bologna, via Ranzani 1, I-40127 Bologna, Italy}

\maketitle
\begin{abstract}
The thermal stability of a weakly magnetized, rotating, stratified,
optically thin plasma is studied by means of linear-perturbation
analysis.  We derive dispersion relations and criteria for stability
against axisymmetric perturbations that generalize previous results on
either non-rotating or unmagnetized fluids. The implications for the
hot atmospheres of galaxies and galaxy clusters are discussed.
\end{abstract}

\section{Introduction}
\label{sec:intro}

Galaxies and galaxy clusters are embedded in hot atmospheres of
virial-temperature gas. The evolution of these systems depends
crucially on whether these atmospheres are subject to thermal
instability \citep{Par53,Fie65,Def70}, so that condensations of cold
gas can form and grow as a consequence of radiative cooling. Motivated
by this astrophysical question, much effort has been devoted to
studying the thermal stability of a stratified, optically thin plasma
by means of both linear-perturbation analysis \citep*[][hereafter
  \citetalias{Nip10}]{Mal87,Loe90,Bal91,Bin09,Bal08,Bal10,Nip10} and
numerical hydrodynamical simulations
\citep*[][]{Kau09,Jou12,McC12,Sha12}.  The thermal-stability
properties of the fluid are influenced by different physical
processes. For instance, unmagnetized media behave differently from
even weakly magnetized media. Similarly, the stability properties
depend significantly on whether the fluid rotates and on the specific
rotation law. While the effect of magnetic fields is accounted for in
several of the aforementioned works, less attention has been paid to
the thermal stability of a rotating stratified plasma.  The effect of
rotation on the thermal stability, which was discussed in simple cases
by \citet{Fie65} and \cite{Def70}, has been studied in detail
analytically by \citetalias{Nip10}, but only for unmagnetized media
(see, e.g., \citealt{DEr98} and \citealt{Li12}, for a numerical
approach).

The rotational properties of the hot gas of galaxies and galaxy
clusters are almost unconstrained observationally, because of the
relatively poor energy resolution of current X-ray instruments. There
are theoretical reasons to expect that rotation is especially
important for the coronae of disc galaxies \citep{Mar11}, but some
models predict significant rotation also for the hot atmospheres of
elliptical galaxies \citep[e.g.][]{Bri09} and galaxy clusters
\citep[e.g.][]{Lau09}.  In any case, even if not dynamically dominant,
rotation influences the thermal-stability properties of the fluid
\citepalias{Nip10}, as it happens for a subthermal magnetic field
\citep{Bal10}.  There is observational evidence of the presence of
magnetic fields in both the intracluster medium \citep{Car02} and the
gaseous halos of galaxies \citep{Jan12}, though the field geometry is
still poorly constrained.  Notwithstanding these uncertainties on the
detailed kinematical and magnetic properties of these gaseous systems,
it is clear that allowing for the presence of both rotation and
magnetic fields would represent an important step forward in the
attempt to understand the thermal stability of the hot atmospheres of
galaxies and galaxy clusters.

Here we focus on the problem of the linear stability of a rotating,
weakly magnetized plasma against axisymmetric perturbations,
accounting for the effects of stratification (i.e. the fluid is in
equilibrium in an external gravitational field), differential
rotation, radiative cooling and anisotropic thermal conduction.  Our
analysis can be considered a generalization of previous studies on the
linear stability of magnetized fluids. For instance, the non-rotating
case has been studied both in the absence \citep{Qua08,Kun11} and in
the presence \citep{Bal08,Bal10,Lat12} of radiative losses. The linear
stability of a rotating magnetized plasma, in the absence of radiative
cooling, has been widely studied in the astrophysical literature, with
specific applications to accretion
discs~\citep[e.g.][]{BalH91,BalH92,Urp98,Kim00,Bra06,Isl12,Sal12} and
rotating stars~\citep[e.g.][]{Fri69,Pit85,Men04,Mas11}.  The
stability-analysis techniques developed in these works, as well as in
studies of rotating unmagnetized media \citepalias[see][and references
  therein]{Nip10}, can be applied to the more general problem
addressed here.

It is worth spending a few words on the structure of the magnetic
field, which, as already mentioned, is poorly constrained
observationally. Throughout this work we will assume that, at least
from the point of view of the perturbation, the magnetic field is
ordered. This does not necessarily imply that the magnetic field is
ordered over scales comparable to the size of the system: the field
may as well be tangled, but with coherence length substantially larger
than the size of the perturbation. If, instead, the coherence length
of the field is comparable to or smaller than the size of the
perturbation, the system is better described by an unmagnetized model
(such as that of ~\citetalias{Nip10}) in which thermal conduction is
suppressed by some factor, accounting for the effect of the tangled
magnetic field (see, e.g., \citealt{Bin09}, for a discussion).

The paper is organized as follows. In Section~\ref{sec:linpert} we
describe the plasma model and we derive the general dispersion
relation for linear axisymmetric perturbations. Stability criteria for
previously studied limiting cases are obtained in
Section~\ref{sec:limiting}.  In Section~\ref{sec:new} we present our
new stability criteria.  In Section~\ref{sec:con} we summarize our
main results and discuss the implications for the hot atmospheres of
galaxies and galaxy clusters.

\section{Linear-perturbation analysis}
\label{sec:linpert}

\subsection{Governing equations}
\label{sec:goveq}

A stratified, rotating, magnetized atmosphere in the presence of
thermal conduction and radiative cooling is governed by the following
magnetohydrodynamics equations:
\begin{eqnarray}
&&{\de \rho \over \de t}+\grad\cdot(\rho\vv)=0,\label{eq:mass}\\
&&\rho\left[{\partial \vv \over \partial t}+\left(\vv\cdot\grad\right) \vv \right]=-\grad \left(p + {B^2 \over 8\pi}\right) -\rho\grad\Phi + {1 \over 4\pi}(\Bv \cdot \nabla)\Bv,\\
&&{\de \Bv \over \de t} - \rot (\vv \times \Bv) = 0, \label{eq:indu}\\
&&{p\over \gamma-1}\left[{\partial \over \partial  t}+\vv\cdot\grad\right] \ln (p \rho^{-\gamma})=
-\grad\cdot \Qv -\rho\L,\label{eq:ene}
\end{eqnarray}
supplemented by the condition that the magnetic field $\Bv$ is
solenoidal ($\div \Bv = 0$).  Here $\rho$, $p$, $T$ and $\vv$ are,
respectively, the density, pressure, temperature and velocity fields
of the fluid, $B\equiv\vert\Bv\vert$, $\Phi$ is the external
gravitational potential (we neglect self-gravity), $\gamma$ is the
adiabatic index,  $\Qv$ is the conductive heat flux, and
$\L=\L(T,\rho)$ is the radiative energy loss per unit mass of
fluid. In a dilute magnetized plasma heat is significantly transported
by electrons only along magnetic force lines
\citep[see][]{Bra65}.  In other words, in the presence of a
  magnetic field the thermal conduction is anisotropic, so the
  conductive heat flux is given by
\begin{equation}
\label{eq:heatflux}
\Qv = -\frac{\chi \Bv\left(\Bv\cdot\nabla\right)T}{ B^2},
\end{equation}
where  $\chi$ is the Spitzer
electron conductivity that can be expressed as
\begin{equation}
\label{eq:chi}
\chi \equiv \kappa T^{5/2},
\end{equation}
 with $\kappa\simeq{1.84\times10^{-5}(\ln{\Lambda})}^{-1} \erg
 \sminusone \cmminusone \kelvin^{-7/2}$, and $\ln\Lambda$ is the
 Coulomb logarithm \citep[][]{Spi62}. In the following we neglect the
 weak temperature and density dependence of $\ln\Lambda$, assuming
 that $\kappa$ is a constant, so $\chi=\chi(T)\propto T^{5/2}$.  In
 the unmagnetized case thermal conduction is isotropic, i.e. the heat
 flux is
\begin{equation}
\label{eq:scalartc}
\Qv = -\kappa T^{5/2} \nabla T.
\end{equation}
Thermal conduction in the presence of a magnetic field is therefore
reduced in directions that are not parallel to the field lines, and it is
null in the direction orthogonal to the field. It follows that the
effect of a tangled magnetic field is, in general, a suppression of
heat conduction: in a first approximation the presence of a tangled
magnetic field can be modeled as an unmagnetized medium with a scalar
thermal conduction (equation~\ref{eq:scalartc}), but with $\kappa$
reduced by some factor \citep[][]{Bin81}. However, if the coherence
length of the tangled field is substantially larger than the size of
the perturbation, the anisotropy of the conductivity becomes crucial
for the stability properties of the magnetized medium
\citep{Bal01,Qua08}. In the following we will focus on the latter case
and we will consider the anisotropic heat flux as given in
equation~(\ref{eq:heatflux}).

 In cylindrical coordinates $(R,\phi,z)$, neglecting all
  derivatives with respect to $\phi$ (because we will consider only
  axisymmetric unperturbed fields and disturbances), the governing
  equations~(\ref{eq:mass}-\ref{eq:ene}) read
\begin{eqnarray}
&&\ds \frac{\de \rho}{\de t} + \frac{1}{R}\frac{\de R\rho \vR}{\de R} + \frac{\de \rho \vz}{\de z} = 0 \label{eq:model_cyl1}, \\
&&\ds \frac{\de \vR}{\de t} + \vR\frac{\de \vR}{\de R}  + \vz\frac{\de \vR}{\de z} - \frac{\vphi^2}{R} = -\frac{1}{\rho}\frac{\de p}{\de R} - \frac{\de \Phi}{\de R} + \frac{1}{4\pi\rho}\left( \BR\frac{\de \BR}{\de R}  + \Bz\frac{\de \BR}{\de z} - \frac{\Bphi^2}{R}\right) - \frac{1}{8\pi\rho}\frac{\de B^2}{\de R},\\
&&\ds \frac{\de \vphi}{\de t} + \vR\frac{\de \vphi}{\de R} + \vz\frac{\de \vphi}{\de z} + \frac{\vR \vphi}{R}= \frac{1}{4\pi\rho}\left( \BR\frac{\de \Bphi}{\de R}  + \Bz\frac{\de \Bphi}{\de z} + \frac{\BR \Bphi}{R}\right), \\
&&\ds \frac{\de \vz}{\de t} + \vR\frac{\de \vz}{\de R}  + \vz\frac{\de \vz}{\de z} = -\frac{1}{\rho}\frac{\de p}{\de z} - \frac{\de \Phi}{\de z} + \frac{1}{4\pi\rho}\left( \BR\frac{\de \Bz}{\de R}  + \Bz\frac{\de \Bz}{\de z}\right) - \frac{1}{8\pi\rho}\frac{\de B^2}{\de z},\\
\label{eq:indR}
&&\ds \frac{\de \BR}{\de t} = \BR\frac{\de \vR}{\de R}  + \Bz\frac{\de \vR}{\de z} - \vR\frac{\de \BR}{\de R}  - \vz\frac{\de \BR}{\de z} - (\div \vv)\BR,\\
\label{eq:indphi}
&&\ds \frac{\de \Bphi}{\de t} = \BR\frac{\de \vphi}{\de R}  + \Bz\frac{\de \vphi}{\de z} + \frac{\Bphi \vR}{R} - \vR\frac{\de \Bphi}{\de R}  - \vz\frac{\de \Bphi}{\de z} - \frac{\vphi \BR}{R} - (\div \vv)\Bphi,\\
\label{eq:indz}
&&\ds \frac{\de \Bz}{\de t} = \BR\frac{\de \vz}{\de R}  + \Bz\frac{\de \vz}{\de z} - \vR\frac{\de \Bz}{\de R}  - \vz\frac{\de \Bz}{\de z} - (\div \vv)\Bz,\\
&&\ds \frac{p}{\gamma -1}\left(\frac{\de}{\de t} + \vR\frac{\de}{\de R} + \vz\frac{\de}{\de z}\right) \ln (p\rho^{-\gamma}) = 
\frac{1}{R}\frac{\de}{\de R}\left[R\frac{\chi(T) \BR (\Bv\cdot\nabla)T}{B^2} \right] 
+ \frac{\de}{\de z} \left[\frac{\chi(T) \Bz (\Bv\cdot\nabla)T}{B^2}\right] 
- \rho\L(T,\rho),
\label{eq:model_cyl2}
\end{eqnarray}
where we have used $\div \Bv = 0$ in writing the three components
(equations \ref{eq:indR}-\ref{eq:indz}) of the induction equation
(\ref{eq:indu}).

\subsection{The unperturbed plasma}
\label{sec:unpert}

The unperturbed system is described by time-independent axisymmetric
pressure $\pzero$, density $\rhozero$, temperature $\Tzero$, velocity
$\vvzero = (\vzeroR, \vzerophi, \vzeroz)$ and magnetic field $\Bvzero
= (\BzeroR, \Bzerophi, \Bzeroz)$ satisfying equations
(\ref{eq:model_cyl1}-\ref{eq:model_cyl2}) with vanishing partial
derivatives with respect to $t$, under the assumption that the plasma
is weakly magnetized, in the sense that the parameter $\beta\equiv
8\pi \pzero/\Bzero^2\gg 1$, where $\Bzero\equiv\vert\Bvzero\vert$: in
other words, the magnetic field is subthermal and dynamically
unimportant. Formally, such a steady-state configuration requires
that, in the unperturbed system, cooling is perfectly balanced by heat
conduction, which appears like an artificial and unrealistic
assumption. However, the steady-state solution can be interpreted more
broadly as describing a ``quasi-stationary'' state, which does not
evolve significantly over the timescales of interest and is close to
hydrostatic and thermal equilibrium. So, globally, the cooling time of
the system is assumed to be much longer than the dynamical time, which
is clearly the case for the hot atmospheres of galaxies and galaxy
clusters.

The unperturbed fluid is allowed to rotate differentially with angular
velocity $\Omega(R,z) \equiv \vzerophi(R,z)/R$ depending on both $R$
and $z$. Without loss of generality we choose our azimuthal coordinate
$\phi$ so that $\Omega\geq0$. We assume that there is no meridional
circulation in the background fluid.  In principle the velocity
components $\vzeroR$ and $\vzeroz$ could be non-null, even in the
absence of meridional circulation, because a time-independent subsonic
inflow of gas can occur if cooling is not perfectly balanced by
thermal conduction (see, e.g., \citetalias{Nip10}).  However, for
simplicity, in the present investigation we limit ourselves to the
case $\vzeroR=\vzeroz=0$.  It must be noted that the Poincar\'e-Wavre
theorem \citep[][]{Tas78}, which holds for unmagnetized fluids,
applies also to our model of magnetized plasma, as long as the
magnetic field is dynamically unimportant in the unperturbed
configuration ($\beta\gg 1$). Therefore, the fluid is baroclinic
[i.e. $p=p(\rho,T)$] in the general case in which
$\Omega=\Omega(R,z)$, and is barotropic [i.e. $p=p(\rho)$] only in the
particular case in which the angular velocity is constant on cylinders
[$\Omega=\Omega(R)$].

 A useful relation among unperturbed quantities is the vorticity
 equation, which is derived from the momentum equations and, under the
 above hypotheses, can be written as
\begin{equation}
\label{eq:vorticity}
R\frac{\de \Omega^2}{\de z}=\frac{1}{\rhozero\Tzero}\left(\frac{\de
  \Tzero}{\de z}\frac{\de \pzero}{\de R}-\frac{\de \Tzero}{\de
  R}\frac{\de \pzero}{\de z}\right).
\end{equation}
As we have assumed that there are no motions in the meridional plane
($\vzeroR=\vzeroz=0$), {\it in the hypothesis that all components of
  the background magnetic fields are time-independent} the induction
equation implies that the unperturbed system satisfies Ferraro's
isorotation law \citep{Fer37}
\begin{equation}
\label{eq:ferraro}
\Bvzero\cdot\nabla\Omega=0,
\end{equation}
i.e. the angular velocity is constant along field lines (see
equation~\ref{eq:indphi}). It has been noted \citep{BalH91} that
condition~(\ref{eq:ferraro}) might be too restrictive when the
magnetic field is weak. If $\Bvzero\cdot\nabla\Omega\neq0$ the
azimuthal component $\Bzerophi$ of the magnetic field varies secularly
(see equation~\ref{eq:indphi}), but, provided that the magnetic field
remains subthermal, the time dependence of $\Bzerophi$ does not imply
the time dependence of any of the other unperturbed fields.  So it is
possible to consider a more general unperturbed weak-field
configuration, in which condition~(\ref{eq:ferraro}) is not satisfied,
$\Bzerophi$ depends on time, while all the other unperturbed
quantities (including $\BzeroR$ and $\Bzeroz$) are time-independent:
the results of the perturbation analysis are still valid in this case,
provided the dispersion relation does not depend on $\Bzerophi$. It
turns out that this is not the case for the most general dispersion
relation derived in the present work (see
Section~\ref{sec:dispersion}), so we will limit ourselves to the case
$\Bvzero\cdot\nabla\Omega=0$.

\subsection{Dispersion relation for axisymmetric perturbations}
\label{sec:dispersion}

We describe here the linear-perturbation analysis of the
weakly magnetized, stratified, dissipative fluid governed by the
equations reported in Section~\ref{sec:goveq}, assuming a background
configuration as described in Section~\ref{sec:unpert}.  In practice,
we linearize the system \eqref{eq:model_cyl1}-\eqref{eq:model_cyl2} by
using axisymmetric Eulerian perturbations of the form $\ds \fzero +
f\e^{-\i\omega t + \i\kR R + \i\kz z}$, where $\fzero$ is the
unperturbed quantity, $|f|\ll|\fzero|$, $\omega$ is the perturbation
frequency, and $\kR$ and $\kz$ are, respectively, the radial and
vertical components of the perturbation wavevector. The linear
analysis is intended to be taken locally in the plasma, in the sense
that the perturbation wavelength is much shorter than the
characteristic scalelengths of the unperturbed system. We further
assume that the perturbation frequency is much lower than the
sound-wave frequency, so we work in the Boussinesq approximation to
exclude a priori the (stable) modes describing the sound waves.  In
linearizing the thermal conduction term, it is useful to note that the
linear perturbation of the heat flux, in the Wentzel-Kramers-Brillouin
(WKB) approximation, is
\begin{equation}
\label{eq:dq}
\Qv =
-\chi(\Tzero) \left[
\bvzero(\bvzero\cdot\nabla)T 
+ \bvzero(\bv\cdot\nabla)\Tzero 
+\bv(\bvzero \cdot\nabla)\Tzero 
-2\bvzero(\bvzero\cdot\bv)(\bvzero\cdot\nabla)\Tzero
\right],
\end{equation}
where we have introduced the dimensionless vectors $\bv\equiv
\Bv/\Bzero$ and $\bvzero\equiv \Bvzero/\Bzero$.  From the above
expression, writing explicitly the axisymmetric perturbation, we
obtain the thermal-conduction term of the linearized energy equation:
\begin{equation}
\label{eq:conduction_linearized}
-\div\Qv 
= \chi(\Tzero) \Tzero \left[ 
-(\kv\cdot\bvzero)^2{T \over \Tzero} 
+ \i(\kv\cdot\bvzero)(\nabla\ln \Tzero \cdot \bv) 
+ \i (\nabla\ln \Tzero \cdot \bvzero)(\kv\cdot\bv) 
-2\i(\bvzero\cdot \kv)(\nabla\ln \Tzero \cdot \bvzero)(\bvzero\cdot \bv)
\right],
\end{equation}
where $\kv=(\kR,0,\kz)$, because $\kphi=0$ for axisymmetric
disturbances.  The first term in the right-hand side of equation
\eqref{eq:conduction_linearized} is similar to what is obtained in the
case of unmagnetized heat flux (equation~\ref{eq:scalartc}), but
accounts for the relative orientation of the displacement and the
unperturbed magnetic field.  The second term leads to the so-called
\emph{magnetothermal instability} (MTI; \citealt{Bal01}). The third
and fourth terms lead to the so-called \emph{heat-flux driven buoyant
  instability} (HBI; \citealt{Qua08}): we recall that these latter two
terms are null if one makes the assumption of isothermal unperturbed
magnetic field lines ($\nabla \Tzero\cdot \bvzero=0$), which implies
that there is no heat flux in the unperturbed medium. In the present
framework, in which the gas is allowed to cool, we are interested in
the more general case $\nabla \Tzero\cdot \bvzero\neq0$, the
underlying assumption being that the timescale of the background heat
flux is long as compared to the local dynamical time. It must be noted
that, even when radiative cooling is negligible, the assumption of
isothermal field lines is not necessary: formally, the unperturbed
system can be in steady state also when $\nabla \Tzero\cdot \bvzero
\neq0$, provided that the divergence of the heat flux vanishes (see
\citealt{Qua08}, for a discussion).

Linearizing the system of partial differential equations
(\ref{eq:model_cyl1}-\ref{eq:model_cyl2}), assuming axisymmetric
perturbations in the WKB approximation, we get the following system of
linear equations:
\begin{eqnarray}
&&\ds \i \kR \vR + \i \kz \vz= 0, \label{eq:linear_axis_con}\\
&&\ds -\i \omega\vR \rhozero + \i \kR p - 2\Omega \vphi \rhozero - \czero^2 \ApR \rho  - \frac{\i \kz \Bzeroz}{4\pi} \BR + \frac{\i \kR}{4\pi}(\Bzerophi \Bphi + \Bzeroz \Bz)= 0, \\
&&\ds -\i \omega \vphi \rhozero + (\Omega + \OmegaR)\vR \rhozero + \Omegaz \vz \rhozero - \frac{\i(\kv \cdot \Bvzero)}{4\pi} \Bphi = 0, \\
&&\ds -\i \omega\vz \rhozero + \i \kz p - \czero^2 \Apz\rho - \frac{\i \kR \BzeroR}{4\pi} \Bz + \frac{\i \kz}{4\pi}(\Bzerophi \Bphi + \BzeroR \BR)= 0, \\
&&\ds -\i \omega\BR - \i (\kv \cdot \Bvzero) \vR = 0, \\
&&\ds -\i \omega\Bphi - \i (\kv \cdot \Bvzero) \vphi - (\OmegaR - \Omega)\BR - \Omegaz \Bz= 0, \\
&&\ds -\i \omega\Bz - \i (\kv \cdot \Bvzero) \vz = 0, \\
&&\ds {1\over\gamma}\left[\i \gamma \omega\frac{\rho}{\rhozero} +
    \vR (\ApR - \gamma \ArhoR) + \vz(\Apz - \gamma \Arhoz)\right] =\nonumber \\
&&\ds \qquad = 
 -\omegad {T \over \Tzero} 
+ \i  \left({\gamma-1\over\gamma}\right)
{\chi(\Tzero) \Tzero \over \pzero}
\left[ 
(\kv\cdot\bvzero)\nabla\ln \Tzero
+(\nabla\ln \Tzero\cdot\bvzero) \kv
-2(\kv \cdot \bvzero)(\nabla\ln \Tzero \cdot \bvzero)\bvzero
\right]  \cdot\bv,
\label{eq:linear_axis_en}
\end{eqnarray}
where $\czero^2\equiv\pzero/\rhozero$ is the isothermal sound speed
squared; the quantities $\ApR\equiv(\pd \pzero/\pd R)/\pzero$
[$\ArhoR\equiv(\pd \rhozero/\pd R)/\rhozero$] and $\Apz\equiv(\partial
\pzero/\pd z) /\pzero$ [$\Arhoz\equiv(\partial \rhozero/\pd z)
  /\rhozero$] are the inverse of the pressure (density) scale-length
and scale-height, respectively.  In analogy with \citetalias{Nip10} we
have defined
\begin{equation}
\omegad\equiv
\begin{cases} 
\omegac + \omegath & \text{if  $\Bzero=0$,}
\\
\omegaca + \omegath & \text{if $\Bzero\neq 0$,}
\end{cases}
\label{eq:omegad}
\end{equation}
where
\begin{equation}
\omegac\equiv\left({\gamma-1\over\gamma}\right){k^2 \chi(\Tzero) \Tzero\over \pzero}\label{eq:omegac}
\end{equation}
is the isotropic thermal-conduction frequency,
  \begin{equation}
  \label{eq:omegaca}
  \omegaca \equiv \frac{(\kv \cdot \bvzero)^2}{k^2}\omegac
  \end{equation}
is the anisotropic thermal-conduction frequency, and
\begin{equation}
\omegath \equiv
-\left({\gamma-1\over\gamma}\right)\frac{\rhozero}{\pzero}
\left[\L(\rhozero,\Tzero)+\rhozero \Lrho(\rhozero,\Tzero) 
-\Tzero \LT(\rhozero,\Tzero) \right]
\label{eq:omegath}
\end{equation}
is the thermal-instability frequency with $\Lrho\equiv {\de
  \L}/{\de\rho}$ and $\LT\equiv {\de \L}/{\de T}$;
$\OmegaR\equiv\partial(\Omega R)/\pd R$ and $\Omegaz\equiv\partial
(\Omega R)/\pd z$ are two frequencies associated with the angular
velocity gradient.  In terms of the defined quantities, the assumption
of short-wavelength perturbations gives
$|\kR|,|\kz|\gg|\ArhoR|,|\Arhoz|,|\Apz|,|\ApR|$, and
$\Omega^2,\OmegaR^2,\Omegaz^2\ll \czero^2 k^2$, while the assumption
of low-frequency perturbations gives $\omega^2 \ll \czero^2 k^2$.  As
implementations of the Boussinesq approximation we neglected the term
$-\i\omega \rho/\rhozero$ in the mass-conservation equation and the
term $-\i\omega p/\pzero$ in the energy equation, and we assumed
$\rhozero T\simeq -\Tzero\rho$ \citepalias[see][]{Nip10}.  We recall
that the magnetic field is considered weak ($\beta\gg1$): in
particular we assumed $\beta$ of the order of $(k/|\ArhoR|)^2$.

The system of linear equations
(\ref{eq:linear_axis_con}-\ref{eq:linear_axis_en}) can be reduced to
the following $5$-th order dispersion relation for $n\equiv-\i\omega$:
\begin{equation}
\label{eq:dispersion_anisotropic}
\begin{array}{c}
\displaystyle n^5 + \omegad n^4 + \left[\omegaBVsq + \omegarotsq + 2 \omegaAsq\right]n^3 + \left[(\omegarotsq + 2 \omegaAsq)\omegad + \omegaAsq\omegacmag \right]n^2 +\\
\displaystyle \quad \,\, + \omegaAsq \left( \omegaAsq 
+ \omegaBVsq + \omegarotsq - 4\Omega^2 \frac{\kz^2}{k^2}+\omegacphisq\right)n + \omegaAsq\left[ \left(\omegaAsq + \omegarotsq - 4\Omega^2\frac{\kz^2}{k^2} \right)\omegad  + \omegaAsq\omegacmag \right] = 0.
\end{array}
\end{equation}
{\it This is the most general dispersion relation derived in the present
work,} describing the evolution of axisymmetric perturbations in a
gravitationally stratified, rotating, plasma subject to radiative
cooling and thermal conduction, in the presence of a weak magnetic
field of arbitrary geometry.  Here, as in \citetalias{Nip10}, we have
introduced the quantity
\begin{equation}
\omegarotsq \equiv -\kzksq{1\over R^3} \Dm({R^4 \Omega^2}),
\label{eq:omegarot}
\end{equation}
which is the square of the frequency associated with differential rotation, and the Brunt-V\"ais\"al\"a
(buoyancy) frequency $\omegaBV$, defined by
\begin{equation}
\omegaBVsq \equiv-\kzksq{\Dm \pzero \over \rhozero\gamma}\Dm \szero,
\label{eq:omegaBV}
\end{equation}
where $\szero \equiv \ln \pzero \rhozero^{-\gamma}$ is the unperturbed
specific entropy and 
\begin{equation}
\Dm \equiv {\kR\over\kz}{\partial\over\partial z}-{\partial\over\partial R}
\label{eq:Dm}
\end{equation}
is a differential operator which takes derivatives along surfaces of
constant wave phase \citep{Bal95}.  In addition we have introduced the
Alfv\'en frequency $\omegaA\equiv \kv\cdot \vvA$, where $\ds\vvA\equiv
{\Bvzero}/{\sqrt{4\pi\rhozero}}$ is the Alfv\'en velocity, and the
following frequencies related to the anisotropy of the thermal
conduction due to the presence of a magnetic field:
\begin{equation}
\label{eq:omegacmag}
\omegacmag \equiv 
-\omegac
\frac{4\pi\pzero}{\Bzero^2}
\kzksq
\frac{\Dm \ln \pzero}{k^2}
\left[
\Dm \ln \Tzero
-2\left(\nabla \ln \Tzero\cdot \bvzero\right)
\left(\frac{\kR}{\kz}{\bzeroz}-\bzeroR\right)
\right],
\end{equation}
which is in general non-null for any geometry of the unperturbed
magnetic field, and $\omegacphi$, defined by
\begin{equation}
\omegacphisq \equiv 
\omegac\Omega
\frac{8\pi \pzero}{\Bzero^2}
\kzksq
\frac{\Dm\ln \pzero}{k^2} 
(\nabla \ln \Tzero\cdot \bvzero)\bzerophi,
\label{eq:omegarot}
\end{equation}
 which can be non-vanishing only if the magnetic field has a
 non-vanishing azimuthal component ($\bzerophi\neq0$).  We note that
 the dispersion relation~(\ref{eq:dispersion_anisotropic}) depends on
 $\Bzerophi$ through the quantities $\bzerophi$, $\bzeroR$ and
 $\bzeroz$, so we will restrict our analysis to unperturbed
 configurations with isorotational magnetic field
 ($\Bvzero\cdot\nabla\Omega=0$), for which $\Bzerophi$ is
 time-independent (see discussion at the end of
 Section~\ref{sec:unpert}).

We recall that, in terms of the quantity $n$ appearing in the
dispersion relation~\eqref{eq:dispersion_anisotropic}, the
perturbation evolves with a time dependence $f(t)\propto \e^{nt}$,
where in general $n\in\mathbb{C}$.  Therefore, stable modes are those
with ${\rm Re}(n)\leq 0$ and unstable modes those with ${\rm Re}(n)>
0$. Among the unstable modes it is useful to distinguish between
purely unstable modes, having ${\rm Im}(n)=0$, in which the
perturbation grows monotonically, and overstable modes, having ${\rm
  Im}(n)\neq0$, in which the perturbation oscillates with
exponentially growing amplitude.  We also remind the reader that,
while $\omegaA^2\geq0$ for all $\kv$, the quantities $\omegarotsq$,
$\omegaBVsq$ and $\omegacphisq$ can be either positive or negative,
also depending on $\kv$. A list of definitions of some relevant
quantities used in the present work is given for reference in
Table~\ref{tab:list}.

\begin{table}
\newcommand\T{\rule{0pt}{4ex}}
\caption{List of definitions.
\label{tab:list}}
\begin{tabular}{l}
\hline
\T$\ApR\equiv\frac{\pd \ln\pzero}{\pd R}$, \qquad$\Apz\equiv\frac{\partial \ln \pzero}{\pd z}$, \qquad$\ArhoR\equiv\frac{\pd \ln\rhozero}{\pd R}$, \qquad $\Arhoz\equiv\frac{\partial \ln\rhozero}{\pd z}$,\\
\T$\czero\equiv \left(\frac{\pzero}{\rhozero}\right)^{1/2}$, \qquad$\Dm\equiv {\frac{\kR}{\kz}}\frac{\partial}{\partial z}-\frac{\partial}{\partial R}$, \qquad $n\equiv-\i\omega$, \qquad$\szero\equiv \ln \pzero \rhozero^{-\gamma}$,\\
\T$\omegaAsq\equiv (\kv\cdot \vvA)^2=\frac{(\kv\cdot{\Bvzero})^2}{4\pi\rhozero}$, \qquad $\omegaBVsq \equiv-\kzksq{\Dm \pzero \over \rhozero\gamma}\Dm \szero$, 
\qquad$\omegac\equiv\left({\gamma-1\over\gamma}\right){k^2 \chi(\Tzero) \Tzero\over \pzero}$, \qquad$\omegaca \equiv \frac{(\kv \cdot \bvzero)^2}{k^2}\omegac$,\\
\T$\omegacmag \equiv 
-\omegac
\frac{4\pi\pzero}{\Bzero^2}
\kzksq
\frac{\Dm \ln \pzero}{k^2}
\left[
\Dm \ln \Tzero
-2\left(\nabla \ln \Tzero\cdot \bvzero\right)
\left(\frac{\kR}{\kz}{\bzeroz}-\bzeroR\right)
\right]$, \qquad $\omegacphisq \equiv 
\omegac\Omega
\frac{8\pi \pzero}{\Bzero^2}
\kzksq
\frac{\Dm\ln \pzero}{k^2} 
(\nabla \ln \Tzero\cdot \bvzero)\bzerophi$,\\
\T$\omegarotsq\equiv  -\kzksq{1\over R^3} \Dm({R^4 \Omega^2})$, \qquad $\omegath \equiv-\left({\gamma-1\over\gamma}\right)\frac{\rhozero}{\pzero}
\left[\L(\rhozero,\Tzero)+\rhozero \Lrho(\rhozero,\Tzero) 
-\Tzero \LT(\rhozero,\Tzero) \right]$,\\ 
\T$\omegad\equiv\omegath+\omegaca$ (if $\Bzero\neq0$), \qquad $\omegad\equiv\omegath+\omegac$ (if $\Bzero=0$),\qquad
$\OmegaR\equiv\frac{\partial(\Omega R)}{\pd R}$,\qquad $\Omegaz\equiv\frac{\partial(\Omega R)}{\pd z}$,\qquad$\alpha\equiv\frac{2\Omega^2
  \rhozero\pzero}{\vert\nabla\pzero\vert^2}$\\ 
\hline
\end{tabular}
\end{table}

\section{Stability criteria in limiting cases: comparison with previous studies}
\label{sec:limiting}

Before investigating the stability conditions for the dispersion
relation~\eqref{eq:dispersion_anisotropic} in its general form, it is
convenient to discuss some simpler cases that have been already
studied in the astrophysical literature.  In particular we analyze
here the limiting cases of the dispersion
relation~\eqref{eq:dispersion_anisotropic} in which at least one among
$\omegaA$, $\omegath$ and $\Omega$ is zero.  In all cases we account
for the effect of thermal conduction on the stability of the plasma.
This preliminary analysis is propaedeutic to Section~\ref{sec:new},
where we present new stability criteria, because it helps simplify the
derivation and the interpretation of the results of the more general
cases discussed there.  In this Section, as well as in
Section~\ref{sec:new}, for a given dispersion relation we obtain
necessary and sufficient conditions for stability following the
approach described in Appendix~\ref{app:RH}, which is based on the
Routh-Hurwitz theorem for the stability of polynomials.

\subsection{Unmagnetized, rotating, radiatively cooling plasma ($\Omega\neq0$, $\omegaA=0$, $\omegath\neq0$)} 
\label{sec:case_N10}

When the Alfv\'en frequency $\omegaA=0$ the dispersion relation
\eqref{eq:dispersion_anisotropic} reduces to
\begin{equation}
\label{eq:dispersion_unmag}
n^3 +n^2\omegad +(\omegaBVsq+\omegarotsq)n+\omegarotsq\omegad=0,
\end{equation}
which was obtained by \citetalias{Nip10} for an unmagnetized rotating
plasma.  We note that the condition $\omegaA=0$ is achieved not only
if the fluid is unmagnetized, but whenever the projection of the
magnetic field onto the wavevector is zero.  For example, the
dispersion relation \eqref{eq:dispersion_unmag} is obtained also for a
system subject to an azimuthal magnetic field ($\Bzerophi\neq0$,
$\BzeroR=\Bzeroz=0$), perturbed with linear axisymmetric disturbances. In this
azimuthal magnetic field configuration thermal conduction is ineffective
($\omegad=\omegath$), in contrast with the unmagnetized case in which
($\omegad=\omegac+\omegath$), but in both cases the dispersion
relation is formally identical to the most general dispersion relation
found in \citetalias{Nip10} for axisymmetric perturbations. Performing
a stability analysis of equation~\eqref{eq:dispersion_unmag} as
described in Appendix~\ref{app:RH}, we obtain the necessary and
sufficient stability criterion
\begin{equation}
\omegad\geq0, \qquad \omegarotsq\geq0,  \qquad\omegaBVsq\geq0,
\end{equation}
as found in \citetalias{Nip10} (see figure 1 in that paper).  For the
above criterion to be satisfied for all wavevectors the following
conditions must hold:
\begin{equation}
\omegath\geq0,\qquad\nabla \pzero\cdot\nabla\szero\leq 0,\quad \frac{\de \Omega}{\de z}=0,\quad \frac{\de (\Omega R^2)}{\de R}\geq0.
\end{equation}
So, we have stability only for barotropic fluids [$\Omega=\Omega(R)$]
satisfying the Field ($\omegath\geq0$), Schwarzschild ($\d \szero/\d
\pzero\leq0$) and Rayleigh $\d(\Omega R^2)/\d R\geq0$ criteria. When
these conditions are not satisfied we can have either monotonically
growing instability or overstability \citepalias[see][]{Nip10}.

\subsection{Magnetized, non-rotating plasma without radiative
  cooling ($\Omega=0$, $\omegaA\neq0$, $\omegath=0$)}
\label{sec:case_Q08}

Here the fluid is magnetized but it does not rotate and there is no
cooling. This is the case considered in Cartesian coordinates by
\cite{Qua08}.  We still work in cylindrical coordinates, but, without
loss of generality, we assume $\Bzerophi=0$ (so $\bzeroR^2+\bzeroz^2=1$).
The dispersion relation \eqref{eq:dispersion_anisotropic} becomes
 \begin{equation}
  \label{eq:dispersion_qua_fact}
  \displaystyle 
\left(n^2 + \omegaAsq\right)
\left[ n^3 + \omegaca n^2 +\left(\omegaAsq + \omegaBVsq \right)n+\omegaAsq\left(\omegaca  + \omegacmag\right)\right]= 0.
  \end{equation}
 The zeroes of the quadratic factor are $n=\pm\i \omegaA$: the two
 oscillatory solutions describing the Alfv\'en waves.  Applying the
 stability analysis (see Appendix \ref{app:RH}) to the cubic factor we
 first get the following two necessary conditions for stability:
  \begin{equation}
  \label{eq:routh_qua2}
\Deltatwo \geq0 \iff  \omegaca\omegaBVsq - \omegacmag\omegaAsq \geq 0,
  \end{equation}
  \begin{equation}
  \label{eq:routh_qua3}
\Deltathree \geq0 \iff  \omegaAsq(\omegaca\omegaBVsq - \omegacmag\omegaAsq)(\omegacmag+\omegaca) \geq 0,
  \end{equation}
because $\Deltaone =\omegaca \geq0$ for all $\kv$. The condition on
$\Deltatwo$ can be rearranged as\footnote{It is useful to notice that
  $\omegaAsq\omegacmag =
  -\omegaca\kzksq\frac{\pzero}{\rhozero}\Dm\ln\pzero \left[ \Dm \ln
    \Tzero -2\left(\nabla \ln \Tzero\cdot \bvzero\right)
    \left(\frac{\kR}{\kz}{\bzeroz}-\bzeroR\right) \right].$}
  \begin{equation}
\label{eq:deltatwoq08}
  \Dm\pzero \left[
    -{1\over\gamma}\Dm\szero + \Dm\ln\Tzero -2\left(\nabla \ln \Tzero\cdot \bvzero\right)
    \left(\frac{\kR}{\kz}\bzeroz-\bzeroR\right)\right]\geq0.
  \end{equation}
Using the identity $ {\Dm\szero} = \gamma\Dm\ln\Tzero
-(\gamma-1)\Dm\ln\pzero$, it is possible to write the condition
\eqref{eq:deltatwoq08} as
  \begin{equation}
  \label{eq:cond_qua2}
  \frac{\gamma-1}{\gamma}\frac{1}{\pzero}\left(\Dm\pzero \right)^2 -2\Dm\pzero\left(\nabla \ln \Tzero\cdot \bvzero\right)
    \left(\frac{\kR}{\kz}\bzeroz-\bzeroR\right)\geq0.
  \end{equation}
We note that in the absence of rotation the fluid is barotropic, so
surfaces of constant density, pressure and temperature coincide (see
Section~\ref{sec:unpert}), and, without loss of generality, we can
assume $\pzero=\pzero(z)$ and $\Tzero=\Tzero(z)$, so that $\Dm=x \d/\d
z$, where $x\equiv \kR/\kz$.  Requiring for stability the validity of
the inequality~\eqref{eq:cond_qua2} $\forall x$, we get the following
two conditions for stability:
\begin{equation}
\bzeroz\bzeroR\frac{\d \Tzero}{\d z}\frac{\d \pzero}{\d z}=0,
\qquad
\frac{\d \pzero}{\d z}\left(\frac{\gamma-1}{\gamma}\frac{\d \ln\pzero}{\d z}-2\bzeroz^2\frac{\d \ln\Tzero}{\d z}\right)\geq0.
\end{equation}
When $\Deltatwo\geq0$ the condition $\Deltathree\geq0$ becomes
$\omegaca+\omegacmag\geq0$, which leads to the following additional
condition:
\begin{equation}
\frac{\d \Tzero}{\d z}\frac{\d \pzero}{\d z}(2\bzeroz^2-1)\geq0.
\end{equation}
Let us now consider the singular cases ($\Deltaone=0$ or $\Deltatwo=0$
or $\Deltathree=0$): it can be shown that the only additional
condition implied by these cases is the Schwarzschild stability
criterion $\omegaBVsq\geq0$, which, in the present case, gives
\begin{equation}
\label{eq:schw}
\frac{\d \pzero}{\d z}\frac{\d \szero}{\d z}\leq0\qquad {\rm i.e.} \qquad
\frac{\d \pzero}{\d z}\left(\frac{\gamma-1}{\gamma}\frac{\d
  \ln\pzero}{\d z}-\frac{\d \ln\Tzero}{\d z}\right)\geq0.
\end{equation}
Summarizing, {\it the necessary and sufficient criterion for stability is}
\begin{equation}
\bzeroz\bzeroR\frac{\d \Tzero}{\d z}\frac{\d \pzero}{\d z}=0,
\qquad
\frac{\d \pzero}{\d z}\left(\frac{\gamma-1}{\gamma}\frac{\d \ln\pzero}{\d z}-2\bzeroz^2\frac{\d \ln\Tzero}{\d z}\right)\geq0,
\end{equation}
\begin{equation}
\frac{\d \Tzero}{\d z}\frac{\d \pzero}{\d z}(2\bzeroz^2-1)\geq0,
\qquad
\frac{\d \pzero}{\d z}\left(\frac{\gamma-1}{\gamma}\frac{\d \ln\pzero}{\d z}-\frac{\d \ln\Tzero}{\d z}\right)\geq0.
\end{equation}
When the unperturbed field lines are isothermal ($\bzeroz=0$,
$\bzeroR=1$), the stability criterion is simply
\begin{equation}
\frac{\d\Tzero}{\d\pzero}\leq 0\quad({\rm for\;stability,\;if}\;\bzeroz=0),
  \end{equation}
where we have used $\Tzero=\Tzero(\pzero)$, because the fluid is
barotropic (in other words the temperature gradient must be opposite
to the pressure gradient).  If this condition is not satisfied we have
the MTI \citep[][]{Bal01}.  When the unperturbed field lines are
not isothermal ($\bzeroz\neq 0$), the stability criterion is 
\begin{equation}
\bzeroR=0, 
\qquad
0\leq\frac{\d\ln\Tzero}{\d\ln\pzero}\leq\frac{\gamma-1}{2\gamma}\quad({\rm for\;stability,\;if}\;\bzeroz\neq0),
  \end{equation}
so we have stability only when temperature gradient, pressure gradient
and magnetic field are parallel, and temperature increases for
increasing pressure, but with logarithmic slope smaller than
$(\gamma-1)/2\gamma$.  If $\bzeroz\neq 0$, but the above stability
conditions are not satisfied,  we have the HBI \citep[][]{Qua08}.

\subsection{Magnetized, non-rotating, radiatively cooling plasma
   ($\Omega=0$, $\omegaA\neq0$, $\omegath\neq0$)}
\label{sec:case_BR10}

In this case the plasma is magnetized and not-rotating, but,
differently from Section~\ref{sec:case_Q08}, we consider the cooling term
in the energy equation \citep[see][]{Bal08,Bal10}. As in
Section~\ref{sec:case_Q08}, the fluid is barotropic and, without loss of
generality, we can consider $\Bzerophi=0$ (i.e. $\bzeroR^2+\bzeroz^2=1$).
The dispersion relation is 
  \begin{equation}
  \label{eq:dispersion_BR_fact}
  \ds 
  \left(n^2 + \omegaAsq\right)
  \left[ n^3 + \omegad n^2 +\left(\omegaAsq + \omegaBVsq \right)n+\omegaAsq\left(\omegad  + \omegacmag\right)\right]= 0.
  \end{equation}
Applying the analysis of Appendix~\ref{app:RH} to the cubic factor, we
first find the necessary conditions for stability
  \begin{equation}
  \label{eq:routh_BR1}
  \Deltaone \geq0 \iff  \omegad \geq 0,
  \end{equation}
  \begin{equation}
  \label{eq:routh_BR2}
  \Deltatwo \geq0 \iff  \omegad\omegaBVsq - \omegacmag\omegaAsq \geq 0,
  \end{equation}
  \begin{equation}
  \label{eq:routh_BR3}
  \Deltathree \geq0 \iff  \omegaAsq(\omegad\omegaBVsq - \omegacmag\omegaAsq)(\omegacmag+\omegad) \geq 0.
  \end{equation}
Imposing the validity of \eqref{eq:routh_BR1} for every wavevector
$\kv$, we find $\omegath>0$, i.e. \citet{Fie65} stability criterion.
Condition \eqref{eq:routh_BR2} can be rewritten as
\begin{equation}
\left(\omegaca + \omegath\right)\omegaBVsq +
\omegaca\kzksq\frac{\pzero}{\rhozero}\Dm\ln\pzero \left[ \Dm \ln
\Tzero -2\left(\nabla \ln \Tzero\cdot \bvzero\right)
\left(\frac{\kR}{\kz}{\bzeroz}-\bzeroR\right) \right] \geq 0,
\end{equation}
which is valid for every wavevector $\kv$ when
\begin{equation}
\label{eq:delta2a}
\omegath\omegaBVsq\geq0,
\end{equation}
\begin{equation}
\label{eq:delta2b}
\omegaBVsq+\frac{\kz^2}{k^2}
\frac{\pzero}{\rhozero}\Dm\ln\pzero
\left[ \Dm \ln \Tzero -2\left(\nabla \ln \Tzero\cdot \bvzero\right)
\left(\frac{\kR}{\kz}{\bzeroz}-\bzeroR\right) \right] \geq0.
\end{equation}
As in Section~\ref{sec:case_Q08}, we can assume $\pzero=\pzero(z)$ and
$\Tzero=\Tzero(z)$, so $\Dm=x \d/\d z$, where $x\equiv \kR/\kz$.
Given that $\omegath>0$, condition~(\ref{eq:delta2a}) reduces to
the classical Schwarzschild criterion $\omegaBVsq\geq0$, which in this
case can be written as condition~(\ref{eq:schw}).  For the
inequality~(\ref{eq:delta2b}) to be satisfied for all $\kv$ we must
have
\begin{equation}
\frac{\d \pzero}{\d z}\frac{\d \Tzero}{\d z}\bzeroz\bzeroR=0 \qquad {\rm and} \qquad \frac{\d \pzero}{\d z}\left(\frac{\gamma-1}{\gamma}\frac{\d \ln\pzero}{\d z}-2\frac{\d \ln\Tzero}{\d z}\bzeroz^2\right)\geq0.
\end{equation}
When $\Deltatwo\geq0$ condition~\eqref{eq:routh_BR3} can be simplified
in $\omegad+\omegacmag\geq0$, which gives
\begin{equation}
\frac{\d \pzero}{\d z}\frac{\d \Tzero}{\d
  z}(1-2\bzeroz^2)x^2+2\frac{\d \pzero}{\d z}\frac{\d \Tzero}{\d z}\bzeroz\bzeroR x\leq0.
\end{equation}
For this to be true for all $x$ we have the additional condition 
\begin{equation}
\frac{\d \pzero}{\d z}\frac{\d \Tzero}{\d z}(1-2\bzeroz^2)\leq0.
\end{equation}
It can be shown that the singular cases ($\Deltaone=0$, $\Deltatwo=0$
and $\Deltathree=0$) do not lead to additional stability criteria.
Summarizing, in the present case {\it the necessary and sufficient
stability criterion is}
\begin{equation}
\omegath>0, \qquad
\frac{\d \pzero}{\d z}\left(\frac{\gamma-1}{\gamma}\frac{\d \ln\pzero}{\d z}-\frac{\d \ln\Tzero}{\d z}\right)\geq0, \qquad
\frac{\d \pzero}{\d z}\frac{\d \Tzero}{\d z}\bzeroz\bzeroR=0,
\end{equation}
\begin{equation}
\frac{\d \pzero}{\d z}\left(\frac{\gamma-1}{\gamma}\frac{\d \ln\pzero}{\d z}-2\frac{\d \ln\Tzero}{\d z}\bzeroz^2\right)\geq0,
\qquad
\frac{\d \pzero}{\d z}\frac{\d \Tzero}{\d z}(1-2\bzeroz^2)\leq0.
\end{equation}
When the unperturbed field lines are isothermal ($\bzeroz=0$) we get:
\begin{equation}
\omegath>0,
\qquad
\frac{\d \Tzero}{\d \pzero}\leq0\qquad({\rm for\;stability,\;if}\;\bzeroz=0);
\end{equation}
when the unperturbed field lines are not isothermal ($\bzeroz\neq 0$),
we get:
\begin{equation}
\omegath>0,
\qquad
\bzeroR=0,
\qquad
0\leq\frac{\d \ln\Tzero}{\d \ln\pzero}\leq \frac{\gamma-1}{2\gamma}\qquad({\rm for\;stability,\;if}\;\bzeroz\neq0),
\end{equation}
in agreement with \citet{Bal10}. So, formally, we can have stability
for specific magnetic field orientations.  However, as well known, the
Field criterion ($\omegath>0$) is typically not satisfied in
astrophysical plasma. When the Field criterion is violated, while the
other stability criteria are met, either overstability or
monotonically growing instability occurs \citep[see][]{Bal10}.

\section{New stability criteria for a weakly magnetized, rotating,
  stratified plasma}
\label{sec:new}

We are now going to analyze the dispersion
relation~\eqref{eq:dispersion_anisotropic}, describing the evolution
of linear perturbations in a stratified, weakly magnetized, rotating
plasma subject to radiative cooling and thermal conduction.  Before
addressing the most general case, which we will discuss in
Section~\ref{sec:coolbphi}, it is convenient to consider separately
cases in which some terms appearing in
equation~\eqref{eq:dispersion_anisotropic} vanish.  In all these
cases, which, as far as we are aware, have not yet been treated in the
astrophysical literature, we account for the fact that the medium is
magnetized ($\omegaA\neq0$), rotating ($\Omega\neq0$) and subject to
thermal conduction along the field lines, and we always allow for the
unperturbed magnetic field lines to be non-isothermal
($\nabla\Tzero\cdot\bvzero\neq0$), but we restrict our analysis to
systems with isorotational unperturbed magnetic field lines
($\bvzero\cdot\nabla\Omega=0$).

\subsection{Plasma in the absence of radiative cooling ($\omegath=0$) with meridional magnetic field ($\Bzerophi=0$)}
\label{sec:nocoolnobphi}

Here we derive stability criteria for a rotating, magnetized plasma in
the absence of radiative cooling.  For simplicity we assume here that
the magnetic field has no azimuthal component, i.e.  $\Bzerophi=0$
(therefore, $\bzeroR^2+\bzeroz^2=1$).  Neglecting radiative losses we get the
dispersion relation
  \begin{equation}
  \begin{array}{c}
  \ds n^5 + \omegaca n^4 + \left(\omegaBVsq + \omegarotsq + 2 \omegaAsq\right)n^3 + \left[(\omegarotsq + 2 \omegaAsq)\omegaca + \omegaAsq\omegacmag \right]n^2 +\\
  \ds \quad \,\, + \omegaAsq \left( \omegaAsq 
  + \omegaBVsq + \omegarotsq - 4\Omega^2 \frac{\kz^2}{k^2}\right)n + \omegaAsq\left[ \left(\omegaAsq + \omegarotsq - 4\Omega^2\frac{\kz^2}{k^2} \right)\omegaca  + \omegaAsq\omegacmag \right] = 0.
  \end{array}
  \label{eq:disprel_nocoolnobphi}
  \end{equation}
Analyzing this dispersion relation as in Appendix~\ref{app:RH} we
first derive the following necessary conditions for stability:
\begin{equation}
\label{eq:routh_B01_2}
\Deltatwo \geq 0 \iff \omegaca\omegaBVsq - \omegacmag\omegaAsq \geq 0,
\end{equation}
\begin{equation}
\label{eq:routh_B01_5}
\Deltafive \geq 0 \iff \omegaca\left(\omegaAsq + \omegarotsq - 4\Omega^2\kzksq\right) + \omegaAsq\omegacmag \geq 0,
\end{equation}
because $\Deltaone$ and $\Deltafour$ are always nonnegative
($\Deltafour=0$ only if $\Deltatwo=0$), and $\Deltathree\geq0$ if
$\Deltatwo\geq0$ and $\Deltafive\geq0$, being
$\Deltathree\propto\Deltatwo\left[\omegaca\left(\omegaAsq +
  \omegarotsq\right) + \omegaAsq\omegacmag\right] $. 
The condition \eqref{eq:routh_B01_2} gives
  \begin{equation}
\Dm\pzero\left[\frac{\gamma-1}{\gamma} \Dm \ln \pzero -2\left(\nabla \ln \Tzero\cdot \bvzero\right)\left(\frac{\kR}{\kz}{\bzeroz}-\bzeroR\right) \right] \geq0.
  \end{equation}
For this to be satisfied for all wavevectors we should have
\begin{equation}
\nabla\pzero\cdot \left[\frac{\gamma-1}{\gamma} \nabla\ln\pzero
    -2\left(\nabla \ln \Tzero\cdot \bvzero\right) \bvzero \right] \geq 0
\qquad
{\rm and}
\qquad
\left(\nabla \Tzero\cdot \bvzero\right)^2\left(\nabla \pzero \times \bvzero\right)^2\leq0.
\end{equation}
Of course the latter condition can be satisfied only with the equality, 
so we can rewrite it as
\begin{equation}
\left(\nabla\Tzero\cdot\bvzero \right)\left(\nabla \pzero \times \bvzero\right)=0.
\end{equation}
Let us now consider the condition~\eqref{eq:routh_B01_5}: the criterion
to have $\Deltafive\geq0$ for all wavevectors is
\begin{equation}
-\nabla\pzero\cdot\nabla\ln\Tzero+\rhozero\frac{\de \Omega^2}{\de \ln R}+2(\nabla\ln \Tzero\cdot\bvzero)(\nabla\pzero\cdot\bvzero)\geq0, 
\end{equation}
\begin{equation}
-\frac{\de \pzero}{\de z}
\left[
\left(\frac{\de \Tzero}{\de z}\frac{\de \Omega^2}{\de R}-
\frac{\de \Tzero}{\de R}\frac{\de \Omega^2}{\de z} \right)
-2\left(\nabla\Tzero\cdot\bvzero \right)
\left(\bzeroz\frac{\de \Omega^2}{\de R}-
\bzeroR\frac{\de \Omega^2}{\de z} \right)\right]
-\frac{\left(\nabla\Tzero\cdot\bvzero \right)^2\left(\nabla\pzero\times\bvzero \right)^2}{R \rhozero \Tzero}
\geq0,
\end{equation}
where we have exploited the vorticity equation~(\ref{eq:vorticity}).
The singular cases to be analyzed separately are $\Deltatwo=0$,
$\Deltathree=0$ and $\Deltafive=0$.  We note that $\Deltatwo=0$ if
either $\Dm\pzero=0$ [i.e. $x=(\de \pzero/\de R)/(\de \pzero/\de z)$,
  if the pressure gradient is non-null] or $x=-\bzeroz/\bzeroR$.  When
$x=(\de \pzero/\de R)/(\de \pzero/\de z)$ the dispersion relation is
\begin{equation}
(n+\omegaca)\left[n^4+(\omegarotsq+2\omegaAsq)n^2+\omegaAsq\left(\omegarotsq+\omegaAsq- 4\Omega^2\frac{\kz^2}{k^2}\right)\right]=0.
\end{equation}
The mode associated with the linear factor ($n=-\omegaca$) is stable
because $\omegaca\geq0$ (conduction damping).  Imposing $n^2<0$ in the
quartic factor, we get
\begin{equation}
\frac{\de \pzero}{\de z}
\left(
\frac{\de \Omega^2}{\de R}\frac{\de \pzero}{\de z}
-\frac{\de \Omega^2}{\de z}\frac{\de \pzero}{\de R}
\right)\geq0.
\end{equation}
When $x=-\bzeroz/\bzeroR$ the dispersion relation is
$n^2+\omegarotsq+\omegaBVsq=0$, so we have stability when
$\omegarotsq+\omegaBVsq\geq0$, which in the considered case
($x=-\bzeroz/\bzeroR$) gives
\begin{equation}
\frac{\gamma-1}{\gamma}\frac{(\bvzero\cdot\nabla\pzero)^2}{\rhozero\pzero}
-\frac{1}{\rhozero}\left(\bvzero\cdot\nabla\pzero\right)\left(\bvzero\cdot\nabla\ln\Tzero\right)
+\bzeroR\left[\frac{\bvzero\cdot\nabla(\Omega^2R^4)}{R^3}\right]\geq0.
\end{equation}
It can be shown that the singular cases $\Deltathree=0$ and
$\Deltafive=0$ do not introduce additional conditions for stability
with respect to those obtained above.

Summarizing, in this case {\it the necessary and sufficient criterion for
stability is}
\begin{equation}
\label{eq:condb01nonisoth1}
\nabla\pzero\cdot \left[\frac{\gamma-1}{\gamma} \nabla\ln\pzero
    -2\left(\nabla \ln \Tzero\cdot \bvzero\right) \bvzero \right] \geq 0,
\qquad
\left(\nabla\Tzero\cdot\bvzero \right)\left(\nabla \pzero \times \bvzero\right)=0,
\end{equation}
\begin{equation}
-\nabla\pzero\cdot\nabla\ln\Tzero+\rhozero\frac{\de \Omega^2}{\de \ln R}+2(\nabla\ln \Tzero\cdot\bvzero)(\nabla\pzero\cdot\bvzero)\geq0, 
\end{equation}
\begin{equation}
-\frac{\de \pzero}{\de z}
\left[
\left(\frac{\de \Tzero}{\de z}\frac{\de \Omega^2}{\de R}-
\frac{\de \Tzero}{\de R}\frac{\de \Omega^2}{\de z} \right)
-2\left(\nabla\Tzero\cdot\bvzero \right)
\left(\bzeroz\frac{\de \Omega^2}{\de R}-
\bzeroR\frac{\de \Omega^2}{\de z} \right)\right]
\geq0,
\end{equation}
\begin{equation}
\label{eq:condb01nonisoth2}
\frac{\de \pzero}{\de z}
\left(
\frac{\de \Omega^2}{\de R}\frac{\de \pzero}{\de z}
-\frac{\de \Omega^2}{\de z}\frac{\de \pzero}{\de R}
\right)\geq0,
\qquad
\frac{\gamma-1}{\gamma}\frac{(\bvzero\cdot\nabla\pzero)^2}{\rhozero\pzero}
-\frac{1}{\rhozero}\left(\bvzero\cdot\nabla\pzero\right)\left(\bvzero\cdot\nabla\ln\Tzero\right)
+4\bzeroR^2\Omega^2\geq0,
\end{equation}
where we have used the isorotation condition
$\bvzero\cdot\nabla\Omega=0$.  It is important to note that the
second of conditions~(\ref{eq:condb01nonisoth1}) requires that, for
stability, we must have either $\nabla \Tzero\cdot\bvzero=0$
(isothermal field lines) or $\nabla\pzero\times\bvzero=0$ (isobaric
surfaces orthogonal to the field lines), so we now consider these two
possibilities.

\subsubsection{Stability criterion when $\nabla \Tzero\cdot\bvzero=0$}

When $\nabla \Tzero\cdot\bvzero=0$, in the current hypothesis of
isorotation ($\bvzero\cdot\nabla\Omega=0$), the isothermal and
isorotational surfaces are parallel to the magnetic field lines, so we
have $\Omega=\Omega(\Tzero)$ and the dispersion relation simplifies
considerably.  Analyzing the dispersion relation in this limit, {\it
  we get the following necessary and sufficient stability criterion}
for a rotating, magnetized medium with isothermal unperturbed field
lines:
\begin{equation}
-\frac{1}{\rhozero}\nabla\pzero\cdot\nabla\ln\Tzero + \frac{\de \Omega^2}{\de\ln R}\geq 0, \qquad \frac{\de \pzero}{\de z}\frac{\de \Tzero}{\de z}\leq 0,
\end{equation}
where we have used the vorticity equation~(\ref{eq:vorticity}).  So a
necessary condition for stability is that the vertical gradients of
temperature and pressure are opposite (we recall that in this case
isothermal and isobaric surfaces do not necessarily coincide).

\subsubsection{Stability criterion when $\nabla\pzero\times\bvzero=0$}

In this case $\nabla\pzero\times\bvzero=0$ and
$\bvzero\cdot\nabla\Omega=0$, so isobaric surfaces are orthogonal to
the field lines and to isorotational surfaces, which implies, for
instance, that
$(\nabla\Tzero\cdot\bvzero)(\nabla\pzero\cdot\bvzero)=\nabla\Tzero\cdot\nabla\pzero$.
 {\it The necessary and sufficient criterion for stability is}
\begin{equation}
\frac{\gamma-1}{\gamma} \vert\nabla\ln\pzero\vert^2 -2\nabla \ln \Tzero\cdot \nabla\ln\pzero \geq 0,
\qquad
\nabla\pzero\cdot\nabla\ln\Tzero+\rhozero\frac{\de \Omega^2}{\de \ln R}\geq0, 
\qquad
\frac{\de \Omega}{\de R}\geq0,
\end{equation}
\begin{equation}
-\frac{\de \pzero}{\de z}
\left[
\left(\frac{\de \Tzero}{\de z}\frac{\de \Omega^2}{\de R}-
\frac{\de \Tzero}{\de R}\frac{\de \Omega^2}{\de z} \right)
-2\left(\nabla\Tzero\cdot\bvzero \right)
\left(\bzeroz\frac{\de \Omega^2}{\de R}-
\bzeroR\frac{\de \Omega^2}{\de z} \right)\right]
\geq0,
\qquad
\frac{\gamma-1}{\gamma}\frac{\vert\nabla\pzero\vert^2}{\pzero}
-\nabla\pzero\cdot\nabla\ln\Tzero
+4\rhozero\bzeroR^2\Omega^2\geq0,
\end{equation}
so a necessary condition for stability is that the angular velocity increases for increasing $R$.

\subsubsection{Stability criteria for a barotropic fluid}

\begin{figure}
\centerline{\psfig{file=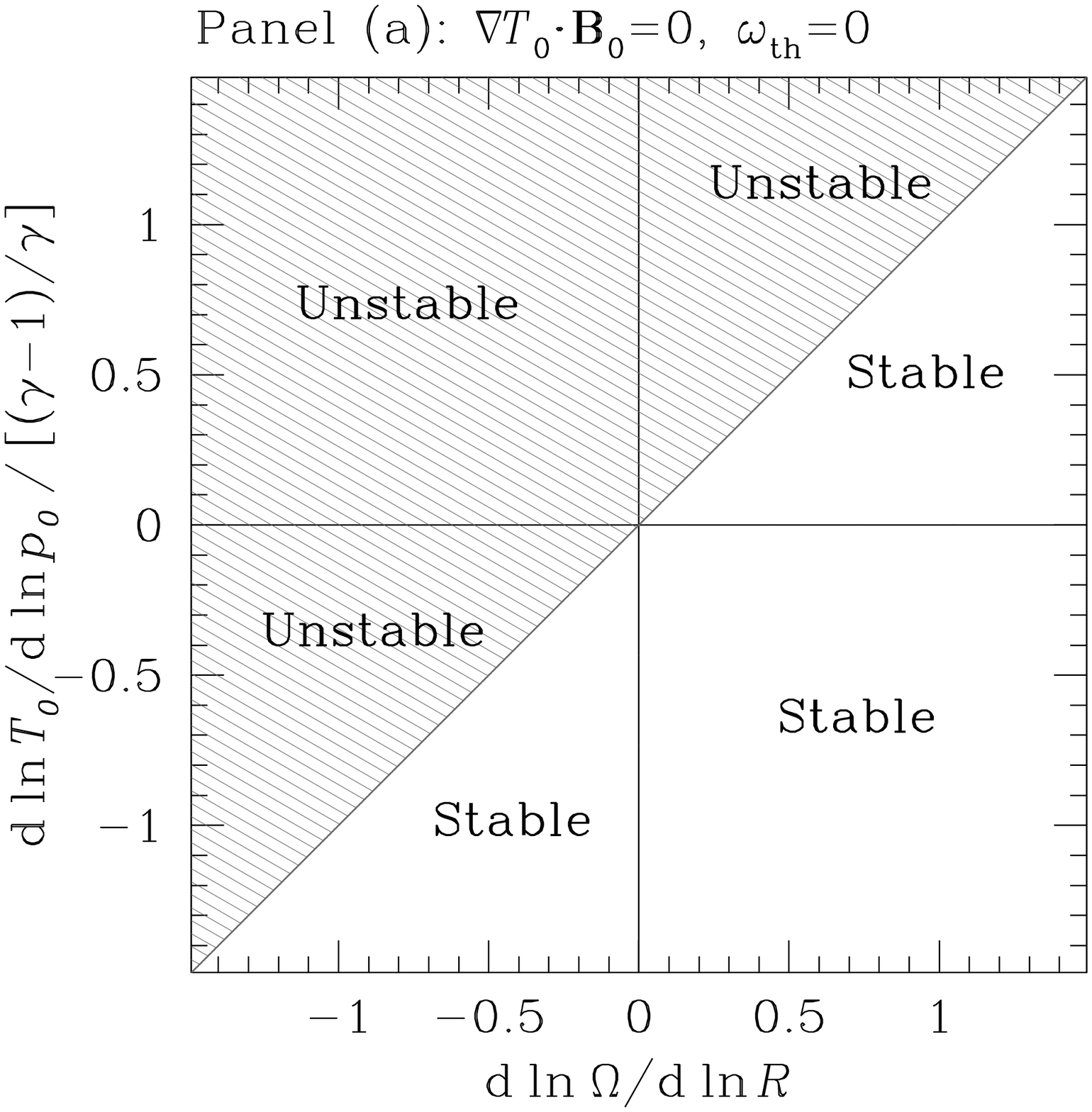,width=0.333\hsize}\psfig{file=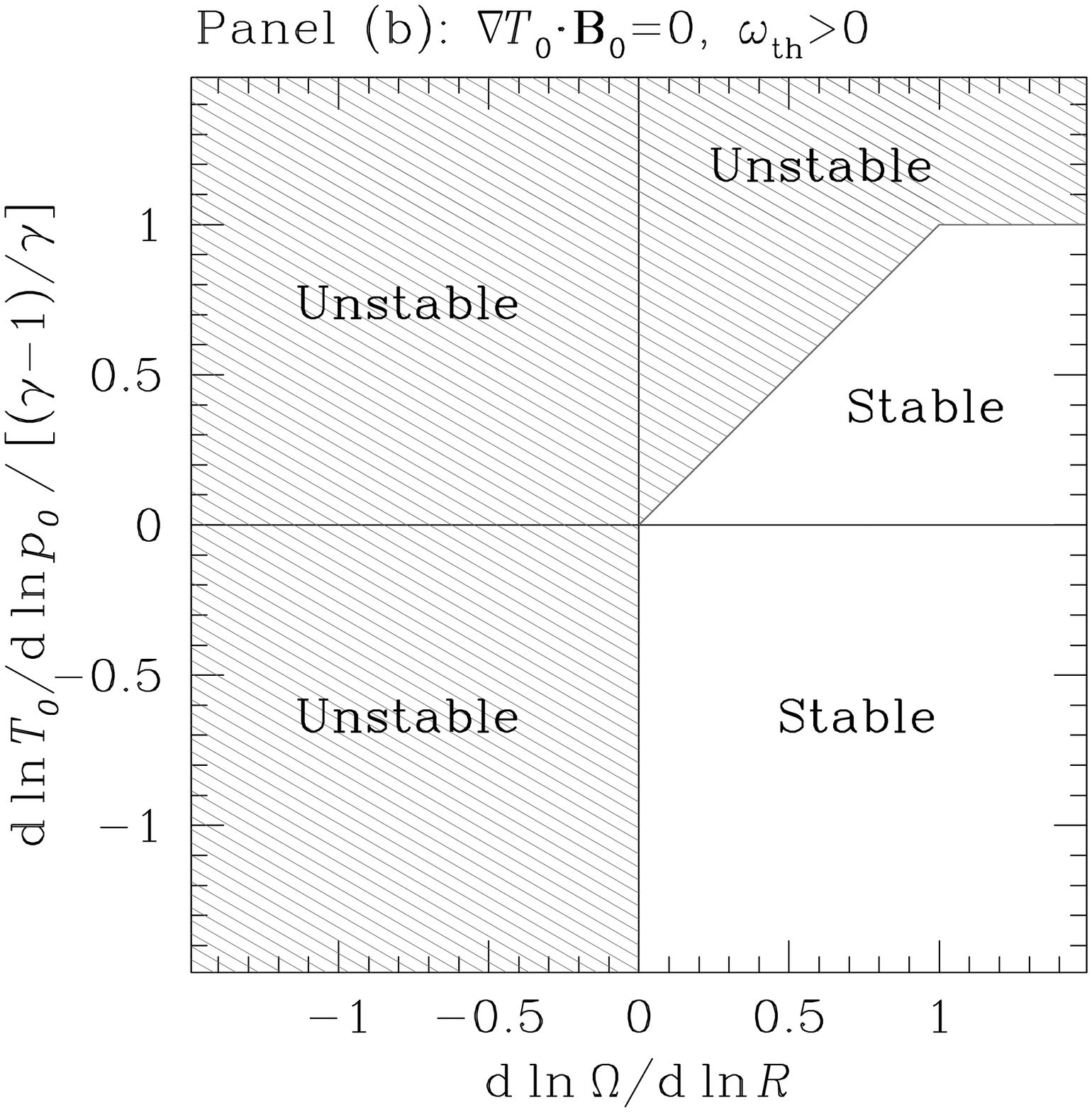,width=0.333\hsize}\psfig{file=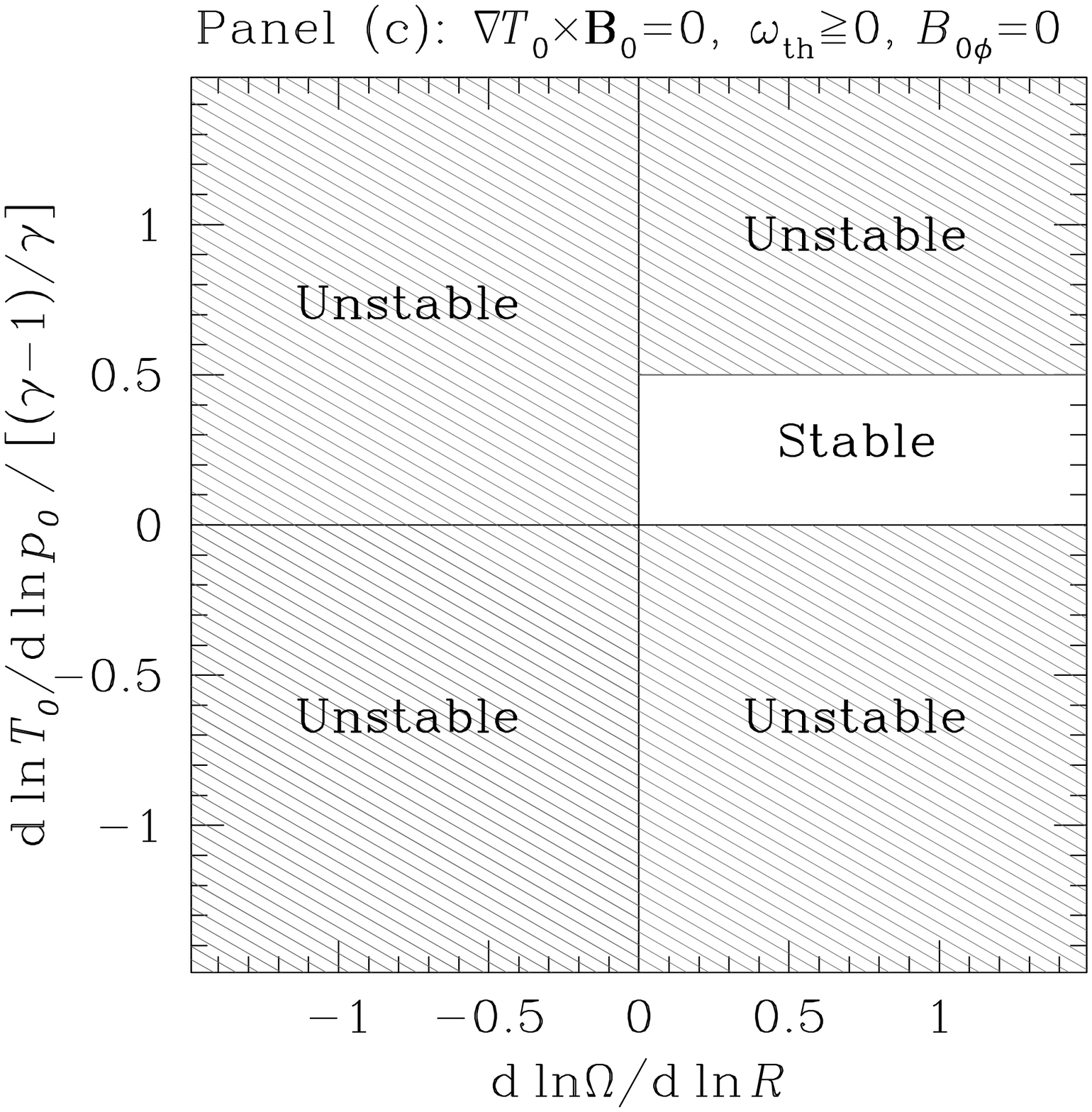,width=0.333\hsize}}
\caption{Stability and instability domains for rotating, barotropic
  [$\Omega=\Omega(R)$] fluids with isorotational unperturbed magnetic
  field lines ($\Bvzero\cdot\nabla\Omega=0$, so $\BzeroR=0$), in the
  space of two parameters: angular-velocity gradient ($x$ axis) and
  gradient of temperature with respect to pressure ($y$ axis). {\it
    Panel (a)} shows the case of a plasma with isothermal unperturbed
  magnetic field lines ($\nabla\Tzero\cdot\Bvzero=0$, so $\de
  \Tzero/\de z=0$) in the absence of radiative cooling ($\omegath=0$;
  see Sections~\ref{sec:nocoolnobphi} and \ref{sec:nocoolbphi}).  {\it
    Panel (b)} shows the case of a radiatively cooling, Field-stable
  ($\omegath>0$) plasma with isothermal unperturbed magnetic field
  lines (see Sections~\ref{sec:coolnobphi} and \ref{sec:coolbphi}).
  {\it Panel (c)} shows the case of a plasma with purely meridional
  unperturbed magnetic field ($\Bzerophi=0$), with field lines
  orthogonal to the isothermal surfaces
  ($\nabla\Tzero\times\Bvzero=0$, so $\de \Tzero/\de R=0$), either in
  the absence of cooling ($\omegath=0$; see
  Section~\ref{sec:nocoolnobphi}) or in the presence of cooling, but
  with $\omegath>0$ (see Section~\ref{sec:coolnobphi}).  In panels (a)
  and (b) $\d\ln\Omega/\d\ln R$ is normalized to
  $(\gamma-1)\gamma^{-1}\alpha^{-1}$, where $\alpha=2\Omega^2
  \rhozero\pzero/\vert\nabla\pzero\vert^2$, so the diagonal lines
  represent $\d\ln\Tzero/\d\ln \pzero=\alpha\,\d \ln\Omega/\d\ln R$. }
\label{fig:omegartp}
\end{figure}

While in the general case of a baroclinic fluid the stability
conditions (\ref{eq:condb01nonisoth1}-\ref{eq:condb01nonisoth2}) are
quite involved, in the special case of a barotropic fluid the
stability criterion becomes considerably simpler and can be easily
shown graphically.  In the barotropic case, the necessary condition
$\left(\nabla\Tzero\cdot\bvzero \right)\left(\nabla \pzero \times
\bvzero\right)=0$ implies that for a configuration to be stable the
magnetic field lines must be either parallel or orthogonal to the
isothermal surfaces [which are also isobaric because
  $\Tzero=\Tzero(\pzero)$]. Note that, when $\Omega=\Omega(R)$ the
isorotation condition~(\ref{eq:ferraro}) implies that the magnetic
field is vertical, so here we must assume $\bzeroR=0$.

When the unperturbed field lines are isothermal
($\nabla\Tzero\cdot\bvzero=0$, which in the current hypotheses implies
$\de\Tzero/\de z=\de\pzero/\de z=\de \Phi/\de z=0$) the stability
criterion reduces to
\begin{equation}
\label{eq:nocoolbaro1}
\frac{\d \ln \Tzero}{\d \ln \pzero}\leq \alpha\frac{\d \ln \Omega}{\d \ln R},\qquad{\rm where}\qquad \alpha\equiv\frac{2\Omega^2
  \rhozero\pzero}{\vert\nabla\pzero\vert^2}
\end{equation}
is a positive dimensionless factor. The corresponding domains of
stability and instability are visualized in
Fig.~\ref{fig:omegartp}(a). Note that, in this special case, there can
be stable configurations with $\d \Omega/\d R<0$, provided the
temperature-pressure gradient is strong and negative. In the limit of
uniform rotation ($\d\Omega/\d R=0$) we recover the condition for
stability against the MTI (see Section~\ref{sec:case_Q08}).

When the unperturbed field lines are parallel to the temperature
gradient ($\nabla\Tzero\times\bvzero=0$, which in the current
hypotheses implies $\de\Tzero/\de R=\de\pzero/\de R=0$ and $\de
\Phi/\de R=\Omega^2R$) the stability conditions
(\ref{eq:condb01nonisoth1}-\ref{eq:condb01nonisoth2}) reduce to
\begin{equation}
0\leq\frac{\d \ln \Tzero}{\d \ln \pzero}\leq\frac{\gamma-1}{2\gamma}
\qquad{\rm and}\qquad
\frac{\d \Omega}{\d R}\geq0,
\end{equation}
which are visualized in Fig.~\ref{fig:omegartp}(c).  So, a barotropic
fluid with magnetic field lines orthogonal to the isothermal surfaces
is stable if and only if the angular velocity increases outwards and
the temperature decreases in the direction of decreasing pressure with
sufficiently shallow gradient.  In other words, the condition of
outward increasing angular velocity must be added to the condition for
stability against the HBI found for non-rotating media (see
Section~\ref{sec:case_Q08}).  We recall that in typical astrophysical
systems the angular velocity decreases outwards: the fact that the
condition ${\d \Omega}/{\d R}\geq0$ is violated is at the heart of the
magnetorotational instability (MRI; \citealt{BalH91,BalH92}), which is
believed to be one of the key mechanisms at work in accretion discs.

\subsection{Plasma in the absence of radiative cooling ($\omegath=0$) with non-vanishing 
azimuthal magnetic field component ($\Bzerophi\neq0$)}
\label{sec:nocoolbphi}

We now discuss the case in which the plasma does not cool, but the
unperturbed magnetic field has a non-vanishing azimuthal component
($\Bzerophi\neq0$). A similar analysis was carried out by
\citet{Bal01}, who assumed isothermal unperturbed magnetic field lines
($\nabla \Tzero\cdot\bvzero=0$): we generalize Balbus' result,
allowing for non-isothermality of the background field lines.  When
$\nabla\Tzero\cdot\bvzero\neq0$ and $\Bzerophi\neq0$, but
$\omegath=0$, the dispersion relation is
\begin{equation}
\label{eq:dispersion_nocoolbphi}
\begin{array}{c}
\displaystyle n^5 + \omegaca n^4 + \left[\omegaBVsq + \omegarotsq + 2 \omegaAsq\right]n^3 + \left[(\omegarotsq + 2 \omegaAsq)\omegaca + \omegaAsq\omegacmag \right]n^2 +\\
\displaystyle \quad \,\, + \omegaAsq \left( \omegaAsq 
+ \omegaBVsq + \omegarotsq - 4\Omega^2 \frac{\kz^2}{k^2}+\omegacphisq\right)n + \omegaAsq\left[ \left(\omegaAsq + \omegarotsq - 4\Omega^2\frac{\kz^2}{k^2} \right)\omegaca  + \omegaAsq\omegacmag \right] = 0,
\end{array}
\end{equation}
which is obtained from equation~\eqref{eq:dispersion_anisotropic},
substituting $\omegad$ with $\omegaca$ (because $\omegath=0$), and
differs from equation~(\ref{eq:disprel_nocoolnobphi}) for the presence of a term
proportional to $\omegacphi^2\propto(\nabla
\Tzero\cdot\bvzero)^2\Bzerophi^2\neq0$.  For stability, we first
impose, following Appendix~\ref{app:RH}, that all the Hurwitz
determinants are nonnegative. We get:
  \begin{equation}
  \label{eq:d2bphi}
  \Deltatwo \geq0 \iff  \omegaca\omegaBVsq - \omegacmag\omegaAsq \geq 0,
  \end{equation}
  \begin{equation}
  \label{eq:d3bphi}
  \Deltathree \geq0 \iff  \Deltatwo\left(\omegaca\omegaAsq+\omegacmag\omegaAsq +\omegaca\omegarotsq\right)-\omegacphisq\omegacasq\omegaAsq\geq 0,
  \end{equation}
  \begin{equation}
  \label{eq:d4bphi}
  \Deltafour \geq0 \iff  
\omegaAsq\left[-\omegacasq\omegaAsq\omegacphi^4
+\left(\omegacmag\omegaAsq+\omegaca\omegarotsq\right)\Deltatwo\omegacphi^2
+4\Omega^2\kzksq\Deltatwo^2\right]\geq 0,
  \end{equation}
  \begin{equation}
  \label{eq:d5bphi}
  \Deltafive \geq0 \iff  \omegaAsq\Deltafour\left[\left(\omegaAsq + \omegarotsq
		-4\Omega^2\kzksq\right) \omegaca + \omegacmag\omegaAsq\right] \geq 0,
  \end{equation}
because $\Deltaone \geq0$ always. Let us consider the condition
$\Deltafour\geq0$. This can be written in the following form:
\begin{equation}
-k^2 f(x)+g(x)\geq0,
\end{equation}
where $x=\kR/\kz$ and\footnote{It can be shown that $f$ depends on the
  wavevector $\kv$ only through $x=\kR/\kz$, so it is independent of
  the wavevector modulus $k$.}
$f(x)=\left(\omegaca\omegaAsq\omegacphisq/k^4\right)^2\geq0$ for all
$x$. Necessary condition for the above inequality to be satisfied for
all $\kv$ (i.e. all $x$ and $k$) is $f(x)=0$ for all $x$. In the
current hypotheses, this implies $\bzeroR=\bzeroz=0$, so
$\kv\cdot\bvzero=0$ for all $\kv$. However, if $\bzeroR=\bzeroz=0$
then $\omegacphi=0$, in contrast with our hypothesis, so we must
conclude that there is no stable configuration with
$\nabla\Tzero\cdot\bvzero\neq0$ and $\Bzerophi\neq0$.

In the case in which the unperturbed field lines are isothermal and
$\Bzerophi\neq0$, $\omegacphi=0$ and the dispersion relation is again
given by equation~(\ref{eq:disprel_nocoolnobphi}) with
$\nabla\Tzero\cdot\bvzero=0$ in $\omegacmag$, which is the same as
that derived, under the same hypotheses, by \citet{Bal01}.  Analyzing
this dispersion relation in the same way as in
Section~\ref{sec:nocoolnobphi}, we get the following stability
conditions:
\begin{equation}
\label{eq:nocoolbphi0}
\nabla\Tzero\cdot\bvzero=0,
\end{equation}
\begin{equation}
\label{eq:nocoolbphi1}
-\nabla\pzero\cdot\nabla\ln\Tzero+\rhozero\frac{\de \Omega^2}{\de \ln R}\geq0, 
\qquad
-\frac{\de \pzero}{\de z}
\left(\frac{\de \Tzero}{\de z}\frac{\de \Omega^2}{\de R}-
\frac{\de \Tzero}{\de R}\frac{\de \Omega^2}{\de z} \right)
\geq0,
\end{equation}
\begin{equation}
\label{eq:nocoolbphi2}
\frac{\de \pzero}{\de z}
\left(
\frac{\de \Omega^2}{\de R}\frac{\de \pzero}{\de z}
-\frac{\de \Omega^2}{\de z}\frac{\de \pzero}{\de R}
\right)\geq0,
\qquad
\frac{\gamma-1}{\gamma}\frac{(\bvzero\cdot\nabla\pzero)^2}{\rhozero\pzero}
+\bzeroR\left[\frac{\bvzero\cdot\nabla(\Omega^2R^4)}{R^3}\right]\geq0.
\end{equation}
The inequalities~(\ref{eq:nocoolbphi1}) coincide with those found by
\citet{Bal01}, while the two conditions~(\ref{eq:nocoolbphi2}), which
follow from the analysis of singular cases of the dispersion
relation~(\ref {eq:disprel_nocoolnobphi}), were not reported before
and, as far as we can see, are generally independent\footnote{The
  conditions~(\ref{eq:nocoolbphi2}) must be considered in addition to
  the conditions~(\ref{eq:nocoolbphi1}) only for baroclinic fluids
  [$\Omega=\Omega(R,z)$], because when the fluid is barotropic
  [$\Omega=\Omega(R)$] the two inequalities~(\ref{eq:nocoolbphi2})
  reduce to $\d\Omega/\d R \geq 0$, which is implied by combining the
  two inequalities~(\ref{eq:nocoolbphi1}).}  of the two
conditions~(\ref{eq:nocoolbphi1}).  Specializing to the case of
isorotational unperturbed field lines, i.e. imposing Ferraro's law
(equation~\ref{eq:ferraro}), the conditions
(\ref{eq:nocoolbphi0}-\ref{eq:nocoolbphi2}) reduce to the following
{\it necessary and sufficient stability criterion:}
\begin{equation}
\label{eq:nocoolbphiiso}
\nabla\Tzero\cdot\bvzero=0,\qquad -\nabla\pzero\cdot\nabla\ln\Tzero+\rhozero\frac{\de \Omega^2}{\de \ln R}\geq0, \qquad\frac{\de\pzero}{\de z}\frac{\de\Tzero}{\de z}\leq0,
\end{equation}
where we have used the vorticity equation~(\ref{eq:vorticity}).  In the case of a
barotropic distribution with isorotational and isothermal field (so
$\bzeroR={\de\pzero}/{\de z}=0$, but $\bzerophi\neq0$), the stability
criterion is again condition~(\ref{eq:nocoolbaro1}), which is
represented in Fig.~\ref{fig:omegartp}(a).

\subsection{Radiatively cooling plasma  ($\omegath\neq0$) with meridional magnetic field ($\Bzerophi=0$)}
\label{sec:coolnobphi}

We move now to the case in which the plasma cools radiatively
($\omegath\neq0$): for simplicity we start here analyzing the
dispersion relation
\begin{equation}
\label{eq:dispersion_coolnobphi}
\begin{array}{c}
\displaystyle n^5 + \omegad n^4 + \left[\omegaBVsq + \omegarotsq + 2 \omegaAsq\right]n^3 + \left[(\omegarotsq + 2 \omegaAsq)\omegad + \omegaAsq\omegacmag \right]n^2 +\\
\displaystyle \quad \,\, + \omegaAsq \left( \omegaAsq 
+ \omegaBVsq + \omegarotsq - 4\Omega^2 \frac{\kz^2}{k^2}\right)n + \omegaAsq\left[ \left(\omegaAsq + \omegarotsq - 4\Omega^2\frac{\kz^2}{k^2} \right)\omegad  + \omegaAsq\omegacmag \right] = 0,
\end{array}
\end{equation}
which is obtained from \eqref{eq:dispersion_anisotropic} in the
hypothesis that the unperturbed magnetic field is meridional
($\Bzerophi=0$), so $\omegacphi=0$. 
Performing a stability analysis as described in Appendix~\ref{app:RH},
we get the following necessary conditions for stability:
\begin{equation}
 \label{eq:d1coolnobphi}
  \Deltaone \geq0 \iff  \omegad \geq 0,
  \end{equation}
  \begin{equation}
  \label{eq:d2coolnobphi}
  \Deltatwo \geq0 \iff  \omegad\omegaBVsq - \omegacmag\omegaAsq \geq 0,
  \end{equation}
  \begin{equation}
  \label{eq:d5coolnobphi}
  \Deltafive \geq0 \iff  \left(\omegaAsq + \omegarotsq
		-4\Omega^2\kzksq\right) \omegad + \omegacmag\omegaAsq \geq 0,
  \end{equation}
because $\Deltafour$ is always nonnegative and
$\Deltathree=\Deltatwo\left[\omegad\left(\omegaAsq+\omegarotsq\right)+\omegaAsq\omegacmag\right]\geq0$
if $\Deltaone\geq0$, $\Deltatwo\geq0$ and $\Deltafive\geq0$.  
Imposing the above conditions for all wavevectors leads to the
following criteria. The condition on $\Deltaone$ gives
$\omegath>0$.  The condition on $\Deltatwo$ gives
\begin{equation}
\label{eq:coolnobphibaro}
\nabla\pzero\times\nabla\Tzero=0\quad \left({\rm i.e.}\;\frac{\de \Omega}{\de z}=0\right),\qquad\nabla\pzero\cdot\nabla\szero\leq0,
\end{equation}
\begin{equation}
\left(\nabla\Tzero\cdot\bvzero \right)\left(\nabla \pzero \times \bvzero\right)=0,
\qquad
\nabla\pzero\cdot \left[\frac{\gamma-1}{\gamma} \nabla\ln\pzero
    -2\left(\nabla \ln \Tzero\cdot \bvzero\right) \bvzero \right] \geq 0.
\end{equation}
Exploiting the fact that we must have $\omegath>0$ (Field
criterion) for the condition on $\Deltaone$, the condition on
$\Deltafive$ gives
\begin{equation}
\frac{\de \Omega}{\de R}\geq0,
\qquad
-\nabla\pzero\cdot\nabla\ln\Tzero+\rhozero\frac{\de \Omega^2}{\de \ln R}+2(\nabla\ln \Tzero\cdot\bvzero)(\nabla\pzero\cdot\bvzero)\geq0, 
\end{equation}
\begin{equation}
-\frac{\de \pzero}{\de z}
\left[
\left(\frac{\de \Tzero}{\de z}\frac{\de \Omega^2}{\de R}-
\frac{\de \Tzero}{\de R}\frac{\de \Omega^2}{\de z} \right)
-2\left(\nabla\Tzero\cdot\bvzero \right)
\left(\bzeroz\frac{\de \Omega^2}{\de R}-
\bzeroR\frac{\de \Omega^2}{\de z} \right)\right]
-\frac{\left(\nabla\Tzero\cdot\bvzero \right)^2\left(\nabla\pzero\times\bvzero \right)^2}{R \rhozero \Tzero}
\geq0.
\end{equation}
The singular cases to be treated separately are $\Deltaone=0$,
$\Deltatwo=0$, $\Deltathree=0$, $\Deltafour=0$ and $\Deltafive=0$, but
it can be shown that they do not lead to additional conditions.  In
summary, taking into account that one of the
conditions~(\ref{eq:coolnobphibaro}) states that the fluid must be
barotropic, {\it the necessary and sufficient stability criterion can
  be written as}
\begin{equation}
\omegath>0,
\qquad
\Omega=\Omega(R),
\qquad
\left(\nabla\Tzero\cdot\bvzero \right)\left(\nabla \pzero \times \bvzero\right)=0,
\qquad
\frac{\d \Omega}{\d R}\geq0,
\qquad\nabla\pzero\cdot\nabla\szero\leq0,
\end{equation}
\begin{equation}
\nabla\pzero\cdot \left[\frac{\gamma-1}{\gamma} \nabla\ln\pzero-2\left(\nabla \ln \Tzero\cdot \bvzero\right) \bvzero \right] \geq 0,
\qquad
-\nabla\pzero\cdot\nabla\ln\Tzero+\rhozero\frac{\d \Omega^2}{\d \ln R}+2(\nabla\ln \Tzero\cdot\bvzero)(\nabla\pzero\cdot\bvzero)\geq0, 
\end{equation}
\begin{equation}
-\frac{\de \pzero}{\de z}
\left[
\frac{\de \Tzero}{\de z}
-2\left(\nabla\Tzero\cdot\bvzero \right)
\bzeroz\right]
\geq0.
\end{equation}
The above conditions must be supplemented by the isorotation law
$\bvzero\cdot\nabla\Omega=0$, which for the relevant barotropic case
implies $\bzeroR=0$ (so $\bzeroz=1$). For a barotropic fluid
$\Tzero=\Tzero(\pzero)$, so the necessary condition
$\left(\nabla\Tzero\cdot\bvzero \right)\left(\nabla \pzero \times
\bvzero\right)=0$ means that for a configuration to be stable the
magnetic field lines must be either parallel or orthogonal to the
isothermal (and isobaric) surfaces.  Let us consider first the case in
which the magnetic field lines are isothermal
($\nabla\Tzero\cdot\bvzero=0$, so $\de\Tzero/\de z=\de\pzero/\de z=0$,
because $\bzeroR=0$). Performing the same analysis as above, in this
limit, we end up with the following necessary and sufficient stability
criterion:
\begin{equation}
\label{eq:coolnobphi1}
\omegath>0,
\qquad
\Omega=\Omega(R),
\qquad
\frac{\d \Omega}{\d R}\geq0,
\qquad
\frac{\d \ln \Tzero}{\d \ln \pzero}\leq \frac{\gamma-1}{\gamma},\qquad \frac{\d \ln \Tzero}{\d \ln \pzero}\leq \alpha\frac{\d \ln \Omega}{\d \ln R}\qquad({\rm when}\;\nabla\Tzero\cdot\bvzero=0),
\end{equation}
where $\alpha>0$ is the dimensionless factor introduced in
equation~(\ref{eq:nocoolbaro1}).  These conditions are represented
graphically in Fig.~\ref{fig:omegartp}(b).  When, instead, the
magnetic field lines are orthogonal to the isothermal surfaces
($\nabla\Tzero\times\bvzero=0$, so $\de\Tzero/\de R=\de\pzero/\de
R=0$, because $\bzeroR=0$), the stability criterion is
\begin{equation}
\label{eq:coolnobphi2}
\omegath>0,
\qquad
\Omega=\Omega(R),
\qquad
\frac{\d \Omega}{\d R}\geq0,
\qquad
0\leq \frac{\d \ln\Tzero}{\d \ln\pzero}\leq\frac{\gamma-1}{2\gamma}
\qquad({\rm when}\;\nabla\Tzero\times\bvzero=0),
\end{equation}
which is visualized in Fig.~\ref{fig:omegartp}(c).

\subsection{Radiatively cooling plasma ($\omegath\neq0$) with non-vanishing azimuthal magnetic field component ($\Bzerophi\neq0$)}
\label{sec:coolbphi}

We now discuss the case in which the unperturbed magnetic field has a
non-vanishing azimuthal component ($\Bzerophi\neq0$) and the plasma is
allowed to cool. This is the most general case studied in the present
work, so the dispersion relation is given by
equation~\eqref{eq:dispersion_anisotropic}, under the assumption that
all terms are non-null (in particular, $\omegath\neq0$ in $\omegad$,
$\nabla\Tzero\cdot\bvzero\neq0$ and $\Bzerophi\neq0$, so
$\omegacphi\neq0$).  For stability let us first impose, following
Appendix~\ref{app:RH}, that all the Hurwitz determinants are
nonnegative. We get:
\begin{equation}
 \label{eq:d1bphi}
  \Deltaone \geq0 \iff  \omegad \geq 0,
  \end{equation}
  \begin{equation}
  \label{eq:d2bphi}
  \Deltatwo \geq0 \iff  \omegad\omegaBVsq - \omegacmag\omegaAsq \geq 0,
  \end{equation}
  \begin{equation}
  \label{eq:d3bphi}
  \Deltathree \geq0 \iff  \Deltatwo\left(\omegad\omegaAsq+\omegacmag\omegaAsq +\omegad\omegarotsq\right)-\omegacphisq\omegadsq\omegaAsq\geq 0,
  \end{equation}
  \begin{equation}
  \label{eq:d4bphi}
  \Deltafour \geq0 \iff  
\omegaAsq\left[-\omegadsq\omegaAsq\omegacphi^4
+\left(\omegacmag\omegaAsq+\omegad\omegarotsq\right)\Deltatwo\omegacphi^2
+4\Omega^2\kzksq\Deltatwo^2\right]\geq 0,
  \end{equation}
  \begin{equation}
  \label{eq:d5bphi}
  \Deltafive \geq0 \iff  \omegaAsq\Deltafour\left[\left(\omegaAsq + \omegarotsq
		-4\Omega^2\kzksq\right) \omegad + \omegacmag\omegaAsq\right] \geq 0.
  \end{equation}
As done in Section~\ref{sec:nocoolbphi}, we note that imposing the
condition $\Deltafour\geq0$ for all wavevectors, in the considered
case in which $\bzerophi\neq0$, we get $\bzeroR=\bzeroz=0$. But if
$\bzeroR=\bzeroz=0$ then $\nabla\Tzero\cdot\bvzero=0$ and
$\omegacphi=0$, in contrast with our hypothesis, so we must conclude
that there is no stable configuration with
$\nabla\Tzero\cdot\bvzero\neq0$ and $\Bzerophi\neq0$.

Let us thus consider the case $\nabla\Tzero\cdot\bvzero=0$ and
$\Bzerophi\neq0$: now $\omegacphi=0$, so the dispersion relation is
equation~(\ref{eq:dispersion_coolnobphi}) and the corresponding
stability conditions are
\begin{equation}
\label{eq:coolbphi1}
\nabla\Tzero\cdot\bvzero=0,
\qquad
\omegath>0,
\qquad
\frac{\de \Omega}{\de z}=0,
\qquad
\frac{\de \Omega}{\de R}\geq0,
\end{equation}
\begin{equation}
\label{eq:coolbphi2}
\nabla\pzero\cdot\nabla\szero\leq0,
\qquad
-\nabla\pzero\cdot\nabla\ln\Tzero+\rhozero\frac{\d \Omega^2}{\d \ln R}\geq0, 
\qquad
-\frac{\de \pzero}{\de z}
\left(\frac{\de \Tzero}{\de z}\frac{\de \Omega^2}{\de R}-
\frac{\de \Tzero}{\de R}\frac{\de \Omega^2}{\de z} \right)
\geq0.
\end{equation}
As in Section~\ref{sec:coolnobphi}, a necessary condition for
stability is $\Omega=\Omega(R)$: note that this is a necessary
requirement also in the absence of magnetic fields (see
Section~\ref{sec:case_N10}), but not in the absence of radiative
cooling (see Sections~\ref{sec:nocoolnobphi} and
\ref{sec:nocoolbphi}).  Let us focus on the case of isorotational
unperturbed magnetic field lines: combining the conditions
(\ref{eq:coolbphi1}-\ref{eq:coolbphi2}) with Ferraro's law
(equation~\ref{eq:ferraro}) we get that {\it the necessary and
  sufficient stability criterion} is
\begin{equation}
\label{eq:coolbphi3}
\nabla\Tzero\cdot\bvzero=0,
\qquad
\omegath>0,
\qquad
\Omega=\Omega(R),
\qquad
\frac{\d \Omega}{\d R}\geq0,
\qquad
\frac{\d \ln \Tzero}{\d \ln \pzero}\leq \frac{\gamma-1}{\gamma},
\qquad \frac{\d \ln \Tzero}{\d \ln \pzero}\leq \alpha\frac{\d \ln \Omega}{\d \ln R},
\end{equation}
where $\alpha>0$ is the dimensionless factor introduced in
equation~(\ref{eq:nocoolbaro1}).  The corresponding domains of
stability and instability are shown in Fig.~\ref{fig:omegartp}(b).

\section{Summary and conclusions}
\label{sec:con}

\subsection{Magnetothermal stability criteria for rotating stratified plasmas}

In the attempt to make a step forward in understanding the thermal
stability of the hot atmospheres of galaxies and galaxy clusters, we
have presented new stability criteria for a gravitationally
stratified, rotating, radiatively cooling, weakly magnetized plasma in
the presence of thermal conduction.  We found that for such a medium
to be stable against {\it all} linear axisymmetric disturbances it
must obey several restrictive conditions, which are reported in
Section~\ref{sec:coolnobphi} (for meridional magnetic field;
equations~\ref{eq:coolnobphi1}-\ref{eq:coolnobphi2}) and
Section~\ref{sec:coolbphi} (for magnetic field with non-null azimuthal
component; equations~\ref{eq:coolbphi1}-\ref{eq:coolbphi2}), and are
represented in Fig.~\ref{fig:omegartp}.  Specifically, in all cases
for stability the fluid is required to be Field-stable
($\omegath>0$), barotropic, with outward increasing angular
velocity ($\d \Omega/\d R\geq 0$). Additional requirements concern the
relative orientation of the magnetic field lines and the thermal
gradient (which must be either parallel or orthogonal) and the
gradient of temperature with respect to pressure (which has different
effects, depending on the relative orientation of the field lines and
the thermal gradient). The above conditions can be seen as a
combination and generalization of criteria for stability against the
MRI \citep{BalH91}, the MTI \citep{Bal01}, the HBI \citep{Qua08} and
the radiatively driven overstability of \citet{Bal10}.  As well known,
even just the two conditions $\omegath>0$ and $\d \Omega/\d R>0$ are
typically not satisfied in standard astrophysical conditions, so we
expect always to find at least one axisymmetric mode that is either
monotonically unstable or overstable.  Therefore, at a formal level,
we must conclude that {\it a radiatively cooling, rotating,
  weakly magnetized astrophysical plasma is not expected to be stable
  against axisymmetric perturbations}.  Similar conclusions were
reached for magnetized (but non rotating) fluids by \citet{Bal10} and
for rotating (but unmagnetized) fluids by \citetalias{Nip10}: in both
cases either overstabilities or monotonic growing instabilities were
expected for typical configurations.  In other words, the calculations
of the present paper have shown that the combination of differential
rotation and a weak ordered magnetic field does not have a stabilizing
effect against thermal axisymmetric perturbations.

In addition to the above results on thermal stability, we have also
generalized previous studies of the stability of magnetized media in
the absence of radiative cooling: in particular we have extended the
study of \citet{Bal01}, by allowing for the presence of non-isothermal
unperturbed field lines, and the study of \citet{Qua08}, by including
the effect of differential rotation (Sections~\ref{sec:nocoolnobphi}
and \ref{sec:nocoolbphi}).  The bottom line of such an analysis is
that isothermality of the unperturbed field lines is a necessary
requirement for stability if the magnetic field has non-vanishing
azimuthal component, but, when the magnetic field is meridional,
stable configurations are formally possible also with (non-isothermal)
field lines orthogonal to the isobaric surfaces.  In any case, the
general conclusion is that, {\it even in the absence of cooling,
  either overstability or monotonic instability is expected}, because
some of the conditions for stability are very unlikely to be satisfied
in real systems (e.g., $\d\Omega/\d R\geq0$ in the barotropic
case). Formally the conditions for stability are more restrictive when
cooling is effective: while in the presence of cooling the
distribution must necessarily be barotropic in order to be stable, in
the absence of cooling it is possible, at least in principle, to have
baroclinic distributions stable against axisymmetric
perturbations. However, these configurations are unlikely to be
astrophysically relevant, because of the requirement of outward
increasing angular velocity.  Therefore it appears that {\it radiative
  cooling is not the key factor determining the instability of a
  rotating magnetized plasma.}

\subsection{Implications for the hot atmospheres of galaxies and galaxy clusters}

At a formal level, the above conclusions are the main results of the
present work. A quite different question is what these results imply
from the astrophysical point of view: in particular, what are the
implications for the evolution of the hot atmospheres of galaxies and
galaxy clusters? We recall that the basic underlying question is
whether cool clouds can condense out of a hot, stratified medium as a
consequence of the instability.  To answer this question one must
necessarily go beyond the formal conclusion that linear instability is
expected and try to investigate the nature of the instabilities
(monotonically growing or overstable), their non-linear evolution and
the dependence on the properties of the perturbation.  Unfortunately,
we cannot address quantitatively these points based only on the
presented calculations, but it is worth discussing them at least
qualitatively.

As first pointed out by \citet[][]{Mal87}, an important aspect of the
problem is the study of the unstable modes, and, in particular,
determining the nature of the instabilities: while monotonically
growing instabilities are expected to naturally lead to condensation,
overstable disturbances are half the time overdense and half the time
underdense, with respect to the local background, so they are likely
disrupted by turbulence before condensing significantly
\citep{Bin09,Jou12}. The problem of the nature of the instabilities
found in the present work could in principle be tackled analytically
by studying the sign of the real roots of the dispersion relations,
but, as mentioned in Appendix~\ref{app:RH}, given the high order of
the involved polynomials the calculations are prohibitively cumbersome
(\citealt{Lia09}, and references therein).  Though a full exploration
of the parameter space is impractical, in selected cases the question
of the nature of the thermal instabilities can be addressed either by
solving numerically the dispersion relations derived in this work or
with magnetohydrodynamics numerical simulations, which can also be
used to gain insight on the non-linear evolution of linearly unstable
perturbations.  For the non-rotating case, an attempt in this
direction has been recently made by \citet{McC12}: the results of
their somewhat idealized magnetohydrodynamics simulations suggest that
no significant condensation occurs in the relevant regime in which the
cooling time is much longer than the dynamical time.

It must be noted that all the stability criteria reported in the
present work concern stability against axisymmetric perturbations.  In
principle, configurations that are stable against axisymmetric
perturbations could be unstable when more general (non-axisymmetric)
perturbations are considered. Therefore, from a mathematical point of
view, exploring different kinds of disturbances could just lead to
even more restrictive stability criteria. However, from the
astrophysical point of view, it is clear that an axisymmetric mode is
not necessarily relevant to real disturbances, and a natural question
to ask is whether the instability occurs also for possibly more
relevant non-axisymmetric perturbations.  As well known, in the
presence of differential rotation, it is considerably more complex to
study non-axisymmetric than axisymmetric perturbations
\citep[e.g.][]{Cow51,BalH92}, and such a task is beyond the purpose of
the present paper.  However, it is interesting to note that, {\it at
  least in the unmagnetized case, non-axisymmetric perturbations tend
  to be stabilized by differential rotation}, while this is not the
case for axisymmetric disturbances \citepalias{Nip10}.

Throughout the paper we have assumed that the magnetic field is
ordered over scales larger than the perturbation size. The
above-presented calculations show that, if this is the case, the fluid
is unlikely to stable, at least against axisymmetric disturbances. It
is hard to predict the outcome of such instability, but in any case we
should expect that also the structure of the magnetic field is
affected. If turbulent motions are produced by the instability, the
(weak) magnetic field is likely tangled by turbulence, and the system
might end up in a configuration which is better described by an
unmagnetized medium with suppressed isotropic thermal conductivity
\citep[][]{Bin09}, so that the results of \citetalias{Nip10} should
apply.  However, it is also possible that, as a consequence of the
instability, the fluid rearranges the magnetic field lines in a
special ordered configuration that tends to contrast the instability,
possibly leading to an overstable state \citep[][]{Bal10}. In the
absence of realistic non-linear calculations it appears difficult to
decide which of the two above hypotheses is a better description of
real systems, and of course alternative scenarios are not excluded.
Overall, as far as the astrophysical implications are concerned, the
results of the present work, though not allowing us to draw general
conclusions on whether cool gas clouds can condense spontaneously out
of an equilibrium plasma, strongly suggest that {\it the interplay of
  rotation and magnetic fields is important for the thermal
  instability}, thus encouraging further theoretical and observational
investigations of the magnetic and rotational properties of the hot
atmospheres of galaxies and galaxy clusters.

\section*{Acknowledgements}

We are grateful to Steven Balbus and James Binney for helpful
discussions. C.N. is supported by the MIUR grant PRIN2008.

\appendix
\section{Analysis of the dispersion relations: stability criteria for parametric polynomials}
\label{app:RH}

The dispersion relations obtained from our linear-perturbation
analysis are polynomials in the complex variable $n=-\i\omega$, whose
coefficients are real and generally depend on the wavevector $\kv$.
To obtain the linear-stability criterion we need to find the conditions
under which the dispersion relation is stable for all wavevectors.
From a mathematical point of view, this problem reduces to studying
the stability of families of polynomials of a complex variable with
real parametric coefficients.

We recall here the {\it Routh-Hurwitz theorem}, which is a fundamental
tool for studying the stability of polynomials of a complex variable
\citep[e.g.][]{Gan59}.  Let us consider an $m$-th degree polynomial
\begin{equation}
p(z)\equiv a_0 z^m + a_1 z^{m-1} + \cdots + a_m; \qquad a_0, a_1, \ldots, a_m \in \mathbb{R},\qquad a_0\neq0,
\end{equation}
where $z\in\mathbb{C}$ and $m \geq 1$: $p(z)$ is said to be a {\it
  Hurwitz polynomial}, or simply a {\it stable polynomial}, if all its
(generally complex) roots have negative real part.  Let us associate
with the polynomial $p(z)$ the following {\it Hurwitz determinants}:
\begin{equation}
\Delta_j\equiv \left| \begin{array}{cccccc}
                   a_1 & a_3 & a_5 & a_7 & \cdots & a_{2j-1}\\
		   a_0 & a_2 & a_4 & a_6 & \cdots & a_{2j-2}\\
		   0 & a_1 & a_3 & a_5 & \cdots & a_{2j-3}\\
		   0 & a_0 & a_2 & a_4 & \cdots & a_{2j-4}\\
		   \vdots & \vdots & \vdots & \vdots & \ddots & \vdots \\
		   0 & 0 & 0 & 0 & \cdots & a_j
                  \end{array} \right|, \qquad j=1, \cdots, m,
\end{equation}
where it has been set $a_i = 0$ for $i>m$.  The Routh-Hurwitz theorem
states that when $\Delta_j \neq 0$ for $j=1,\dots,m$, the polynomial
$p(z)$ is stable if and only if all Hurwitz determinants are strictly
positive:
\begin{equation}
\Delta_1 >0, \, \Delta_2 >0, \cdots, \, \Delta_m >0.
\end{equation}
A corollary of the theorem is that, under the same hypotheses, a
necessary condition for $p(z)$ to be stable is that all its real
coefficients be strictly positive; i.e. $a_0, a_1, \ldots, a_m >0$.

It must be stressed that when at least one of the Hurwitz determinants
is null the Routh-Hurwitz theorem does not apply.  In our application
to the dispersion relations the coefficients of the polynomial are
parametric and it is not generally the case that all the Hurwitz
determinants are non-null. Therefore, in order to obtain necessary and
sufficient conditions for stability, we proceed as follows.  We first
find the conditions to have
\begin{equation}
\label{eq:deltageq}
\Delta_1 \geq0, \, \Delta_2\geq0, \cdots, \, \Delta_m \geq0,
\end{equation}
for all wavevectors. At this stage these conditions must be considered
necessary, but not sufficient, because it is not guaranteed that the
singular cases ($\Delta_i=0$ for at least one $i$) are stable.  We
then consider separately the dispersion relations obtained in the
singular cases, thus obtaining additional stability conditions that,
combined with those obtained from~(\ref{eq:deltageq}), give the
necessary and sufficient stability criterion.

When the criterion for stability is not met, one is interested in
determining whether the unstable modes are monotonically growing ($z$
is real and positive) or overstable ($z$ is complex, with positive
real part). Mathematically this requires a complete root
classification of polynomials with real parametric coefficients. This
task, which can be easily accomplished for third-order polynomials
such as those considered in \citetalias{Nip10}, is extremely complex
for the higher-order polynomials we deal with in this work
\citep[e.g.][]{Lia09}. Therefore, in the present study we limit
ourselves to deriving stability criteria and we do not attempt a
classification of the nature of the unstable modes.

\end{document}